%% file: ms.tex
\documentclass[usenatbib]{mn2e}
\usepackage{amsmath,amssymb,graphicx}
\usepackage{longtable,lscape,subfigure}  
\usepackage[usenames, dvipsnames]{color}
\input{def}

\title[X-ray flare from a candidate spiral galaxy]{Unveiling the Nature of an X-ray flare from 3XMM\thanks{Based on observations obtained with {\sc XMM-Newton}, an ESA science mission with instruments and contributions directly funded by ESA Member States and NASA} J014528.9+610729: A candidate spiral galaxy}

\author[Bhatt et al.]{Himali  Bhatt$^{1}$\thanks{mshimali@gmail.com}, Subir Bhattacharyya$^{1}$, Nilay Bhatt$^1$, J. C. Pandey$^2$\\
$^{1}$Astrophysical Sciences Division, Bhabha Atomic Research Centre, Trombay, Mumbai 400 085, India \\
$^{2}$Aryabhatta Research Institute of observational sciencES, Manora Peak, Nainital 263 129, India\\
}

\begin{document}

\date{\today}

\pagerange{\pageref{firstpage}--\pageref{lastpage}} \pubyear{2014}

\maketitle   
\label{firstpage}

\begin{abstract}
We report an X-ray flare from 3XMM J014528.9+610729, serendipitously detected during the observation of the open star cluster NGC 663.
The colour-colour space technique using optical and infrared data reveals the X-ray source as a candidate spiral galaxy. 
The flare shows fast rise and exponential decay shape with a ratio of the peak and the quiescent count rates of $\sim$60 and duration of $\sim$5.4 ks.
The spectrum during the flaring state is well fitted with a combination of thermal ({\sc Apec}) model with a plasma temperature of $\rm{1.3\pm0.1}$ keV 
and non-thermal ({\sc Power-law}) model with power-law index of $\rm{1.9\pm0.2}$.
However, no firm conclusion can be made for the spectrum  during the quiescent state.
The temporal behavior, plasma temperature and spectral evolution during flare suggest that the flare from  3XMM J014528.9+610729 can not be associated with
tidal disruption events.

\end{abstract}


\section{Introduction}
\label{sec:intro}

The  X-ray emission from normal galaxies is mainly associated with  
bright high-mass X-ray binaries (HMXBs),
Supernova remnants (SNRs),
O-type stars and hot gas heated by energy originated in supernova explosions 
\citep{per+02,fab06}.  
The hard X-ray (2-10 keV) emission is dominated by HMXBs, and the soft X-ray (0.3-2.0 keV) emission
is mostly produced by the gas at $kT \sim$ 0.3-0.7 keV \citep{san+11}.
Giant X-ray outbursts with flare peak  to quiescent state flux ratio up to a factor $\sim$200 from non-active galaxies have also been detected 
with extreme X-ray softness, e.g., NGC 5905, RXJ1242-11, RXJ1624+75, RXJ1420+53, RXJ1331-32 
\citep{kom+01}.
Tidal disruption of a star by a supermassive black hole (SMBHs) in the nuclei of galaxies 
is considered as the favoured explanation for these unusual events \citep[e.g.,][]{rees88}.
Based on a  luminous flare seen in 
soft X-rays,  several candidate tidal disruption events (TDE) have been identified so far
\citep[e.g.,][]{rees88,gre+00,esq+07,cap+09,sax+12}.
These events show high peak luminosities (up to $\sim$ $\rm{10^{44} erg~s^{-1}}$), very soft spectra
characterized by thermal emission in energy range 0.04-0.1 keV and the X-ray flux fall over the long-term as $\rm{t^{-5/3}}$ \citep[see][]{kom+02}.
\citet{kom+99} discussed the possibility of such outburst in a non-active galaxy due to accretion disk instability.
A localized instability in an advection dominated disk can lead to such outbursts from a non-active galaxy.
Therefore, X-ray outbursts from non-active galaxies provide important
information on the presence of SMBHs in these
galaxies and the link between active and normal galaxies \citep{kom+01}.

In this paper, we analyzed an X-ray flare detected from an X-ray source during the X-ray observation 
of the open cluster NGC 663 from {\sc XMM-Newton }  \citep{bhat+13, bhat+14}.
This source is given in 3XMM-DR4 catalogue as 3XMM J014528.9+610729, which is the sixth publicly released {\sc XMM-Newton}
 X-ray source catalogue produced by the XMM Survey Science Centre.
We made an attempt to classify this source using  multiwavelength data and it has been argued
that the source is a candidate spiral galaxy.  
We describe the X-ray data reduction procedure and information of the multiwavelength data used in the present study  in \S2.
The identification methods of the X-ray source using multiwavelength data are given in \S3.
In \S4, we present the temporal and spectral analysis of the X-ray data.
Finally, we discussed our results in \S5 and draw the conclusions in \S6.    

\begin{figure*}
\subfigure[]{\includegraphics[width=85mm]{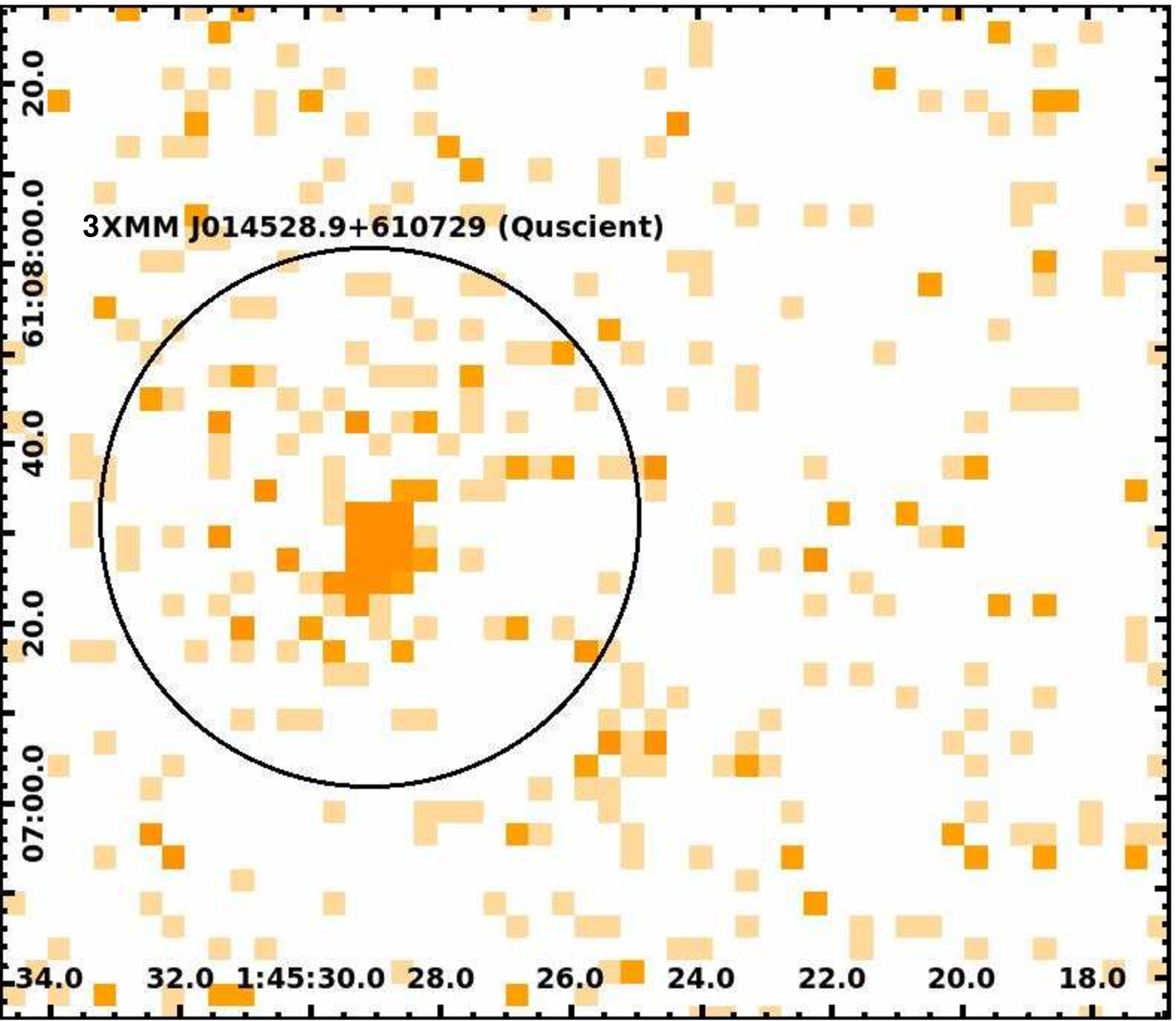}}
\subfigure[]{\includegraphics[width=85mm]{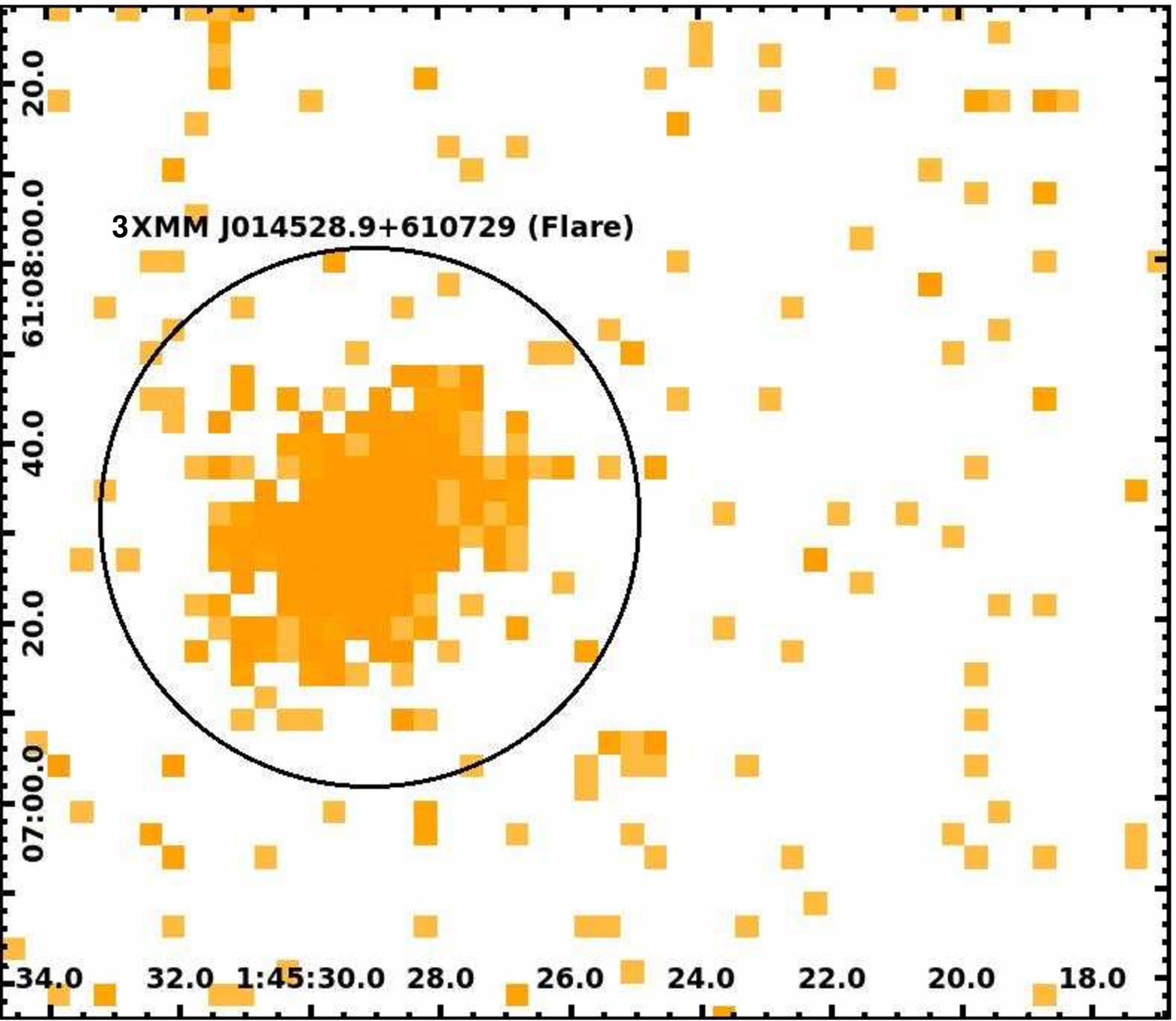}}
\caption{X-ray image of 3XMM J014528.9+610729.(a) before the flaring event and (b) during
the flaring event. X-axis and Y-axis are representing RA(J2000) and DEC(J2000), respectively.}
\label{fig:image}
\end{figure*}

\section{Observations and Data Reduction}
\label{sec:reduc}
3XMM J014528.9+610729 is serendipitously observed by  {\sc XMM-Newton} during the observation 
of the young open cluster NGC 663 on $\rm{14^{th}}$ January 2004 at 22:23:02 UT (53018.93266 MJD) corresponding to
observation identification number 0201160101.
The data obtained in the  {\sc XMM-Newton} observations have been reduced using 
the Science Analysis Software \citep[SAS;][]{gib+04} version 12.0.1.
The standard procedure adopted for the reduction of
European Photon Imaging Camera (EPIC)  and Optical Monitor (OM) 
data are given below. The data from the Reflection Grating Spectrometer \citep[RGS;][]{brin+98,her+01} have not been used in the present study
because the X-ray source is $\sim$9\arcmin~away from the center of the field of view
of RGS\footnote{http://XMM.esac.esa.int/external/XMM\_user\_support/ documentation/uhb/rgs.html} (FOV$\sim$5\arcmin) during observations.

\subsection{EPIC data}
EPIC constitutes the PN CCD  detector \citep{stu+01}, and the twin CCD detectors MOS1 and MOS2 \citep{tur+01}.
EPIC was used in full frame mode during observations with Medium filter for an exposure time of $\sim$ 42 ks. 
Calibrated event files were created using SAS
tasks {\sc epchain} and   {\sc emchain} for PN and MOS detectors, respectively.
The images and lightcurves of the event list 
were extracted using the SAS task {\sc evselect}.
The high background periods were excluded from the observations 
where the total count rate (for single events of energy above 10 keV) in the instruments
exceeds 0.35 and 1.0 $\rm{counts~s^{-1}}$ for the MOS and PN detectors, respectively.
The sums of good time intervals were found to be 32.59 ks, 
33.13 ks and 28.69 ks for PN , MOS1 and MOS2 detectors, respectively.
The detail description of the data reduction procedure is given in \citet{bhat+13}.
Further, we selected single and double pixel events (corresponding to PATTERN $\leq$ 4) for PN and all valid events for MOS (PATTERN = 0--12).
The FLAG$==$0 was then used for selection of events for both PN and MOS cameras.
The resultant events were  used for the extraction of source image, source and background lightcurves and spectra.
The images of the X-ray source 3XMM J014528.9+610729 in the energy band 0.3--10.0 keV
during quiescent and flaring states are shown in Figure~\ref{fig:image}.
The circular region with radius  24$\arcsec$ around the source  3XMM J014528.9+610729 was used for lightcurve and spectrum  extraction.
The background was taken locally from identical (equal area) circular region located on the same CCD where the source was positioned for both PN and MOS detectors.
The background region was selected in this way to avoid inclusion of bad pixels.

\subsection{OM data}
OM is a f/12.7 Ritchey Chretien telescope coaligned with the X-ray telescopes and operating simultaneously 
with them \citep[for details see][]{man+10}. 
The OM was configured in imaging mode, by using  V filter\footnote{http://XMM.esac.esa.int/external/XMM\_user\_support/ documentation/uhb/omfilters.html} ($\lambda_{eff}\sim$ 5430 \AA; $\Delta\lambda\sim$70 \AA) during observations.
Eight exposures were taken with integration time of 1748 s for each of first seven exposures
(S006-S012) and 2798 s for the last one (S013).
The OM covers the central 17$\arcmin \times$ 17$\arcmin$ region of the 
X-ray field of view during all the exposure intervals except for S011 (4$\arcmin \times$17$\arcmin$).    
The images, source lists and magnitudes of the sources were produced by the SAS tool {\sc omichain}. 

\subsection{Multiwavelength archival data}
Multiwavelength data were used for source identification as the spectral class and type of the source has not yet been derived.
The data from the following surveys have been used to classify the source.

\subsubsection{Optical data : SDSS}
The optical data from the Sloan Digital Sky Survey \citep[SDSS;][]{aba+09} have been used in the present study.
The SDSS uses a dedicated wide-field 2.5 m telescope \citep{gun+06} 
located at Apache Point Observatory (APO) near Sacramento Peak in Southern New Mexico. 
The SDSS photometric systems \citep{fuk+96} 
$u^\prime$ (3543\AA; $\delta \lambda\sim$ 567\AA),
$g^\prime$ (4770\AA; $\delta \lambda\sim$ 1387\AA),
$r^\prime$ (6231\AA; $\delta \lambda\sim$ 1373\AA),
$i^\prime$ (7625\AA; $\delta \lambda\sim$ 1526\AA) and 
$z^\prime$ (9134\AA; $\delta \lambda\sim$ 950\AA),
are similar to the AB system \citep{oke+83}.
By matching the position of the X-ray source  3XMM J014528.9+610729  
with the position of the optical sources in SDSS catalog,
the optical source J014528.91+610729.5 is found to be the closest to the X-ray source 
3XMM J014528.9+610729 with an
offset of $\rm{0.003^{\prime\prime}}$.
The magnitudes of the optical source SDSSJ014528.91+610729.5 in SDSS bands are given in Table~\ref{tab:sed}.

\subsubsection{Near-Infrared data: 2MASS}
The Near -Infrared (NIR) data were taken from the Two Micron All Sky Survey \citep[2MASS;][]{cut+03} in 
 $J$ (1.25 $\micron$), $H$ (1.65 $\micron$) and $K_s$ (2.17 $\micron$) bands.
The cross-correlation of the position of X-ray source 3XMM J014528.9+610729 with the 2MASS catalog shows that
the 2MASS source 2MASSJ01452893+6107292 is closest to the 3XMM J014528.9+610729 with an offset of   $\rm{0.324^{\prime\prime}}$,
and its $J$, $H$ and $K_s$ magnitudes are listed in Table~\ref{tab:sed}.    

\subsubsection{Near and Mid Infrared data: WISE}
Wide-field Infrared Survey Explorer \citep[WISE;][]{wri+10} mapped the sky at 3.4, 4.6, 12, and 22 $\micron$ 
(W1, W2, W3, W4) in 2010 with an angular resolution of 6.1$\arcsec$, 6.4$\arcsec$, 6.5$\arcsec$ and 12.0$\arcsec$, respectively.
The magnitudes were taken in W1, W2, W3 and W4 bands from WISE All-Sky Data Release products \citep{cut+12}.
The closest counterpart of 3XMM J014528.9+610729 in WISE catalog is found within an offset of $\rm{0.626^{\prime\prime}}$, namely,
WISEJ014528.93+610728.8 and its magnitudes in WISE bands are tabulated in Table~\ref{tab:sed}. 

\section{Source identification using multiwavelength data}
\label{sec:iden}
The multiwavelength data are required to classify the X-ray source  
into the various source types - stars, galaxies, clusters, and active galactic nuclei (AGN).
We have searched the X-ray source 3XMM J014528.9+610729 into SDSS, 2MASS and WISE sky surveys covering 
wavelength ranging from optical to mid IR.
The values of on-axis angular resolution\footnote{http://XMM.esac.esa.int/external/XMM\_user\_support/ documentation/uhb/onaxisxraypsf.html}  (FWHM on ground)  
are $\rm{6.6^{\prime\prime}}$, $\rm{6.0^{\prime\prime}}$ and
$\rm{4.5^{\prime\prime}}$ for PN, MOS1 and MOS2 detectors, respectively.   
The nearest counterparts of the  X-ray source 3XMM J014528.9+610729
are given in Table~\ref{tab:sed}.
We found only one counterpart of 3XMM J014528.9+610729 in each catalog within $\rm{4.5^{\prime\prime}}$ search radius, which is the best possible 
resolution from {\sc XMM-Newton}. Therefore, all these multiwavelength sources, which are within an offset of 
$\rm{\sim0.7^{\prime\prime}}$, may correspond to the X-ray source 3XMM J014528.9+610729.
Using the multiwavelength  information of 3XMM J014528.9+610729, we opted the following procedure to classify the X-ray source.  

\begin{table*}
\normalsize
\caption{Cross identification of X-ray source 3XMM J014528.9+610729 into  SDSS, 2MASS and WISE sky surveys.} 
\begin{tabular}{llllrl}
\hline
Survey  & Name                 & offset     & Band           & Magnitudes           & Flux$^\dagger$\\
        &                      & ($\arcsec$)&                &  (mag)               & ($\rm{10^{-3}}$ Jy) \\
\hline
SDSS    & J014528.91+610729.5  & 0.003    &  $u^\prime$    & 22.35$\pm$0.32       & 2.71$\pm$ 0.81 \\
        &                      &          &  $g^\prime$    & 19.869$\pm$0.016     & 6.18$\pm$ 0.09  \\	
        &                      &          &  $r^\prime$    & 18.354$\pm$0.008     & 5.32$\pm$ 0.04  \\ 
        &                      &          &  $i^\prime$    & 16.758$\pm$0.005     & 9.48$\pm$ 0.04  \\ 
        &                      &          &  $z^\prime$    & 15.876$\pm$0.006     &10.83$\pm$ 0.06  \\ 
\hline                                                                                    
2MASS  & J01452893+6107292     & 0.324    & J              & 14.413$\pm$0.041     & 8.03$\pm$ 0.30 \\	
       &                        &         & H              & 13.832$\pm$0.044     & 5.94$\pm$ 0.24 \\ 
       &                        &         & $K_s$          & 13.592$\pm$0.045     & 3.86$\pm$ 0.16 \\ 
\hline                                                                                    
WISE   & J014528.93+610728.8   & 0.626    & W1             & 13.377$\pm$0.030     & 1.84$\pm$ 0.05  \\	      
       &                       &          & W2             & 13.257$\pm$0.035     & 1.06$\pm$ 0.03  \\ 
       &                       &          & W3             & 10.133$\pm$0.074     & 3.72$\pm$ 0.25 \\ 
       &                       &          & W4             &  8.90$\pm$0.37       & 2.85$\pm$ 0.98  \\
\hline
\end{tabular}
\newline
$^\dagger$ : These fluxes are extinction corrected (see \S\ref{sec:sed}).
\label{tab:sed}
\end{table*} 

\subsection{Colour-colour diagrams}
\label{sec:cc}
\citet{fan99} simulated  the "fundamental plane" in colour space of 
 ($u^\prime$ - $g^\prime$),  ($g^\prime$ - $r^\prime$), ($r^\prime$ - $i^\prime$) and  ($i^\prime$ - $z^\prime$) for
normal stars, white dwarfs (WD),  halo blue horizontal branch
stars (BHBs) as well as quasars (QSO) and the compact emission-line galaxies (CELGs).
All four kinds of stellar objects ( stars, WD, CELs and BHBs) are distributed basically on the same plane in colour space, but
quasars are located in a different plane.
Therefore, these simulations are very useful to identify QSOs. 
The  ($u^\prime$ - $g^\prime$),  ($g^\prime$ - $r^\prime$), ($r^\prime$ - $i^\prime$) and  ($i^\prime$ - $z^\prime$) colours are
estimated to be 2.45$\pm$0.32 mag, 1.51$\pm$0.05 mag,  1.596$\pm$0.009 mag and  0.0882$\pm$0.008 mag, respectively. 
The X-ray source 3XMM J014528.9+610729 (see Figure A in supplementary material) 
is located far away from the locus of QSOs , but above the locus of stars.
Therefore, we can discard the possibility of the X-ray source 3XMM J014528.9+610729 for being a QSO, however, 
it is very difficult to distinguish between stars and galaxies using optical SDSS data.
Therefore, we have used NIR and MIR colour-colour diagrams to distinguish it from the stars.

Recently, \citet{tu+13} defined a $(J-K_s)$ and $(K_s-W3)$ colour-colour plane
to distinguish
asymptotic giant branch (AGB) stars from normal stars, galaxies and QSOs.
The colour-colour plane provides the  1-, 2-, and 3-$\sigma$ regions of the normal stars,
1- and 2$\sigma$ regions of the galaxies, and the QSOs.
The  $(J-K_s)$  and $(K_s-W3)$ colours of the X-ray source 3XMM J014528.9+610729 are estimated to be 
0.82$\pm$0.06 mag and 3.46$\pm$0.09 mag, respectively.
The  X-ray source 3XMM J014528.9+610729 (see  Figure B in supplementary material)
lies near the region occupied by galaxies, which is outside the  3-$\sigma$  boundary
of the normal stars.

Further using WISE (W1-W2) and (W2-W3) colours, \citet{wri+10} showed the regions 
occupied by  stars, brown dwarfs, elliptical galaxies, spiral galaxies,
startburst galaxies, luminous IR galaxies (LIRGs), 
 low-ionization nuclear emission-line regions (NLERs) galaxies,
ultraluminous infrared galaxies (ULIRGs),
QSOs, Seyferts and obscured AGNs. 
The WISE (W1-W2) and (W2-W3) colours for the X-ray source 3XMM J014528.9+610729 are found to be 
3.124$\pm$0.082 mag and 0.12$\pm$0.05 mag, respectively (see Figure C in supplementary material), and
is located in the
region of spiral galaxies, but near the regions  occupied by LIRGs. 
Therefore, on the basis of multiwavelength colour-colour diagrams, the
X-ray source 3XMM J014528.9+610729 is very likely to be a  spiral galaxy.  

\subsection{Spectral energy distribution}
\label{sec:sed}

\begin{figure*}
\includegraphics[width=100mm,angle=270]{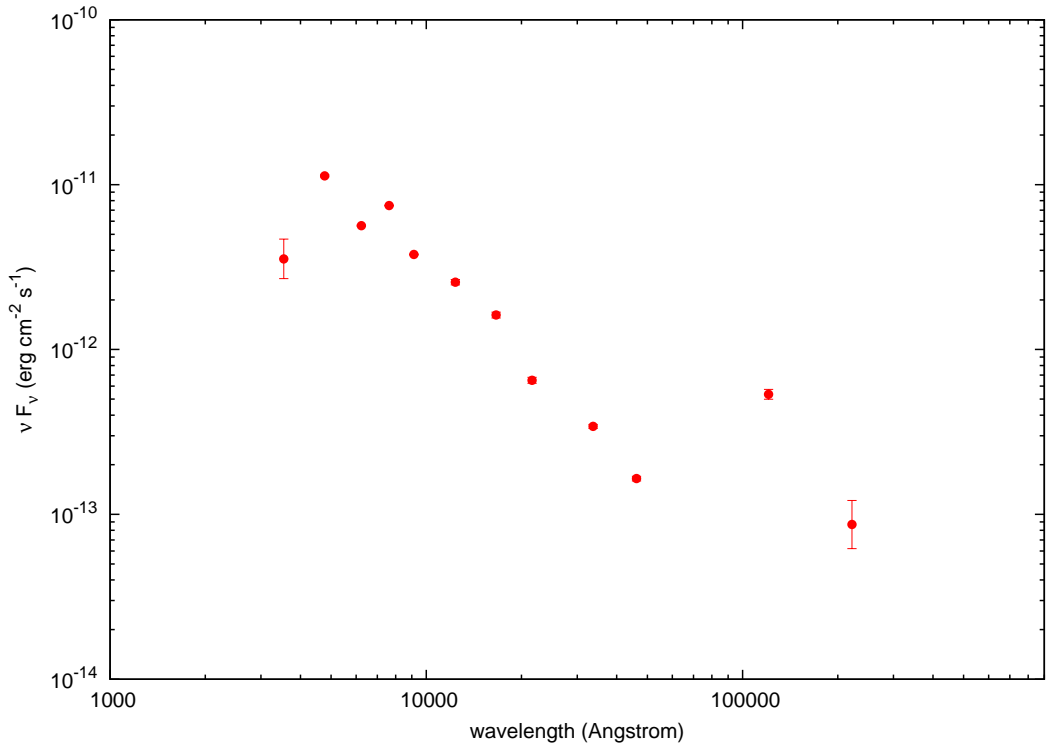}
\caption{ Extinction corrected observed frame spectral energy distribution of 3XMM J014528.9+610729. The  Vertical error bars show 1$\sigma$ uncertainties due to photometric uncertainties on the fluxes.}
\label{fig:sed}
\end{figure*}

The reddening towards the direction of 3XMM J014528.9+610729 in $V$ band is given as  $\rm{A_V}$ =4.522 mag in 
NASA/IPAC Extragalactic Database\footnote{http://ned.ipac.caltech.edu/} using
Galactic extinction from \citet{sch+11}. 
The reddening ($\rm{A_\lambda}$) towards the direction of 3XMM J014528.9+610729 for SDSS wavebands 
 $u^\prime$,  $g^\prime$, $r^\prime$, $i^\prime$ and $z^\prime$ are given as 6.990, 5.447, 3.768, 2.800 and 2.083  mag, respectively.
For 2MASS $J$, $H$ and $K_s$ bands, the $\rm{A_\lambda}$ are given as 1.169, 0.740 and 0.498 mag, respectively.
The reddening in WISE wavebands are derived using the relations given in \citet{gan+11}\footnote{$\rm {A_{W1}= 0.0697 \times A_\lambda}$,
$\rm {A_{W2}= 0.0527 \times A_\lambda}$,
$\rm {A_{W3}= 0.068 \times A_\lambda}$,
$\rm {A_{W4}= 0.0517 \times A_\lambda}$} in W1, W2, W3 and W4 wavebands and
estimated to be 0.315, 0.238, 0.307 and 0.234 mag, respectively.  
The spectral energy distribution (SED) of 3XMM J014528.9+610729 is shown in Figure \ref{fig:sed}.

Using the colour-colour information, we have fitted the SED of the 
X-ray source 3XMM J014528.9+610729 with the templates of different types of galaxies 
using the template fitting procedure given by \citet[referred as Hyperz]{bol+00} and \citet{ass+08,ass+10}.
The Galaxy template from  \citet{ass+08,ass+10} is best fitted with $\chi^2$ of 11330 (dof 11), which implies that
the object is outside the parameter space covered by the models.
Using Hyperz template fitting procedure, the data are best fitted with the spiral galaxy (SB2) template with 
$\chi^2_\nu$ of 457 (dof 11) with a  fitting probability of 0\%.
Therefore, none of the fitting procedure is able to give an acceptable $\chi^2$ using any of the galaxy templates.
Here, we are not able to classify the source based on  the template fitting procedure, and therefore we cannot
determine the redshift of the X-ray source. 

\section{Results}

\subsection{X-ray lightcurves}
\label{sec:lc}
\begin{figure*}
\includegraphics[width=160mm]{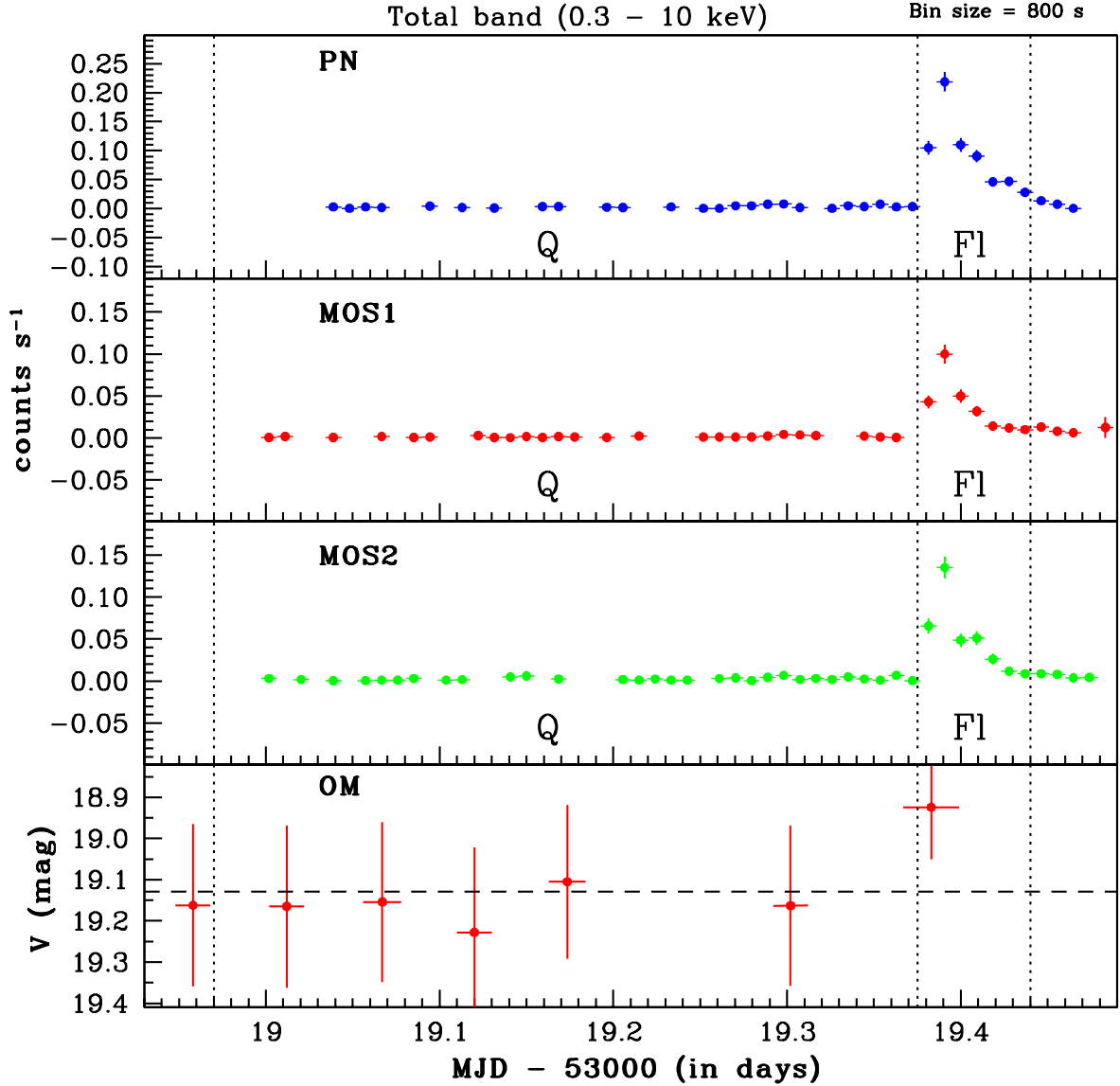}
\caption{Background corrected lightcurves in EPIC (energy band 0.3 - 10 keV) and OM detectors. Quiescent states (Q) and Flaring states (Fl) are marked by dotted line in each panel. }
\label{fig:lightcurves}
\end{figure*}

X-ray lightcurves were extracted using {\sc SAS} task {\sc evselect}. 
The background lightcurves from adjacent source-free regions were also accounted for and
the background scaling factors were calculated using {\sc backscale} task. 
To check the variation in different energy bands, 
the lightcurves were built in three energy bands --  total (T; 0.3-10.0 keV), soft (S; 0.3-2.0 keV) and  hard (H, 2.0-10.0 keV) 
with a time binwidth of 800 s.
The background subtracted lightcurves are shown  in Figure~\ref{fig:lightcurves} 
for the total energy band, and in Figure~\ref{fig:lightcurves_sub} for soft and hard energy bands.
X-ray lightcurves show flaring features where flares are characterized by two or more 
consecutive time bins that constitute a sequence of either rising or falling count rates, 
corresponding to rise and decay phase of the flare. The flare regions and  quiescent regions 
are  marked by "Fl" and "Q"  in Figure~\ref{fig:lightcurves} by dotted lines.

$\chi^2$--test has been performed to estimate the statistical significance of the flare-like variability in the lightcurves and
the $\chi^2$ values with a degree of freedom (dof) are given  in Table~\ref{tab:chi}.
The values of the probabilities of variability ($\rm{P_{var}}$)  in lightcurves for each detector have been calculated 
and are found to be greater than 99.999\%.
Fractional root mean square (rms) variability amplitudes ($\rm F_{var}$) 
are estimated to quantify the amplitude of variability in the X-ray lightcurves. 
The  $\rm F_{var}$ has been defined as \citep{ede+02,ede+90}

\begin{equation}
\rm {F_{var}  = \frac{1}{<X>} \sqrt{S^2 - <\sigma^{2}_{err}>}} 
\label{eq:fvar}
\end{equation}

\begin{equation}
\rm  {\sigma_{F_{var}} = \frac{1}{F_{var}} \sqrt{\frac{1}{2N}}  \frac{S^2}{<X>^2}}
\label{eq:fvarerr}
\end{equation}

\noindent where $S^2$ is the total variance of the lightcurve, $<\sigma^{2}_{err}>$
is the mean error squared, $<X>$ is the mean count rate and  \rm{$\sigma_{F_{var}}$} is the error in $\rm F_{var}$. 
The values of $\rm F_{var}$ and corresponding errors are given in Table~\ref{tab:chi}.

\begin{table*}
\caption{Timing analysis using $\chi^2$-test with the probabilities of rejection of  the null hypothesis ($\rm{P_{var}}$) with dof 
in different energy bands-- Total (T, 0.3-10.0 keV), Soft (S, 0.3-2.0 keV) and Hard (H, 2.0-10.0 keV), for each {\sc Epic} detectors.  
The amplitude of variability in the X-ray lightcurves is defined using $\rm F_{var}$.}    
\begin{tabular}{llllllllll}
\hline
Energy         &                                  \multicolumn{3}{c}{$\chi^2$-Test [$\chi^2$ (dof)] }              &   \multicolumn{3}{c}{Variability amplitude ($\rm F_{var}$) } \\
\cline{2-7}
                &  \multicolumn{1}{c}{PN}     & \multicolumn{1}{c}{MOS1}    & \multicolumn{1}{c}{MOS2}   &    \multicolumn{1}{c}{PN}     & \multicolumn{1}{c}{MOS1}    & \multicolumn{1}{c}{MOS2} \\ 
\hline
T           & 1944(34)& 461(36) & 828(40)                      & 2.09$\pm$0.25 & 2.06$\pm$0.25 & 2.26$\pm$0.25 \\
S           & 1947(32)& 508(29) & 815(39)                      & 2.03$\pm$0.25 &  1.79$\pm$0.24 &  2.26$\pm$0.26\\
H           &   97(31)&  67(27) &  58(26)                      & 1.73$\pm$0.23 &  1.65$\pm$0.25 &  1.32$\pm$0.21\\
\hline
\end{tabular}
\label{tab:chi}
\end{table*} 

In energy band T, the mean count rates during quiescent state 
are estimated to be 0.003$\pm$0.002, 0.002$\pm$0.001 and 0.002$\pm$0.001 counts $\rm s^{-1}$ in
PN, MOS1 and MOS2 detectors, respectively.
The duration of the X-ray flare is $\sim$5.6 ks, with rise and decay times  $\sim$1.6 ks and $\sim$4.0 ks, respectively.
It shows a rapid rise and slower decay in count rates, and the peak count rates at flaring state are found to be nearly 73$\pm$50, 50$\pm$25 and 65$\pm$34 times 
more than that of the quiescent state in PN, MOS1 and MOS2 detectors, respectively.

The X-ray flare from the candidate spiral galaxy 3XMM J014528.9+610729  appears highly asymmetric with fast rise time and long decay.
The decay time scales of the flares are very important to understand the physical mechanism of generation of flares.
The X-ray flares from the quiescent galaxies are mainly associated with the TDEs  and
the flux decay of TDE flares is broadly consistent with
a power law with a slope of $\sim$-5/3 \citep[e.g.,][]{rees88,gre+00,esq+07,cap+09,sax+12}.
However, the X-ray flares from AGNs are having fast rise and exponential decay (FRED) shape \citep[e.g.,][]{mar+99,fos+04}.
Therefore, to understand the behaviour of the flare, we have fitted the count rate $c(t)$ as a function of time $\it t$ during the decay phase 
in lightcurves of PN detector
with power law decay and exponential decay using the following equations, respectively.

\begin{equation}
\rm { c(t) = a \times t^{-\alpha} +b }  
\label{eq:powdecay}
\end{equation}

\noindent where  a and b are constants and $\alpha$ is the power-law index. The decay phase is not well fitted with a {\sc Power-law} model 
as suggested by $\chi^2_{\nu}$ value of  14.6 with  dof 5.  

\begin{equation}
\rm { c(t) = A_0 exp^{-[(t-t_0)/\tau_{d}]} + q   }
\label{eq:expfun}
\end{equation}

\noindent where $\it {t_0}$ is the time of peak count rate, $\it q$ is the count rate in the quiescent
state (0.003 $\rm{counts~s^{-1}}$), $\it \tau_d$ is the decay time of the flare and $\it A_0$ is the count rate at flare peak. 
The best-fit values of $\it \tau_d$ is estimated to be 1707$\pm$144 s with $\chi^2_{\nu}$ of 1.48 with dof 6.
Thus, the flare is well fitted with the FRED shape. 
 
\subsection{Optical lightcurve}
\begin{figure*}
\vspace{2cm}
\includegraphics[width=85mm]{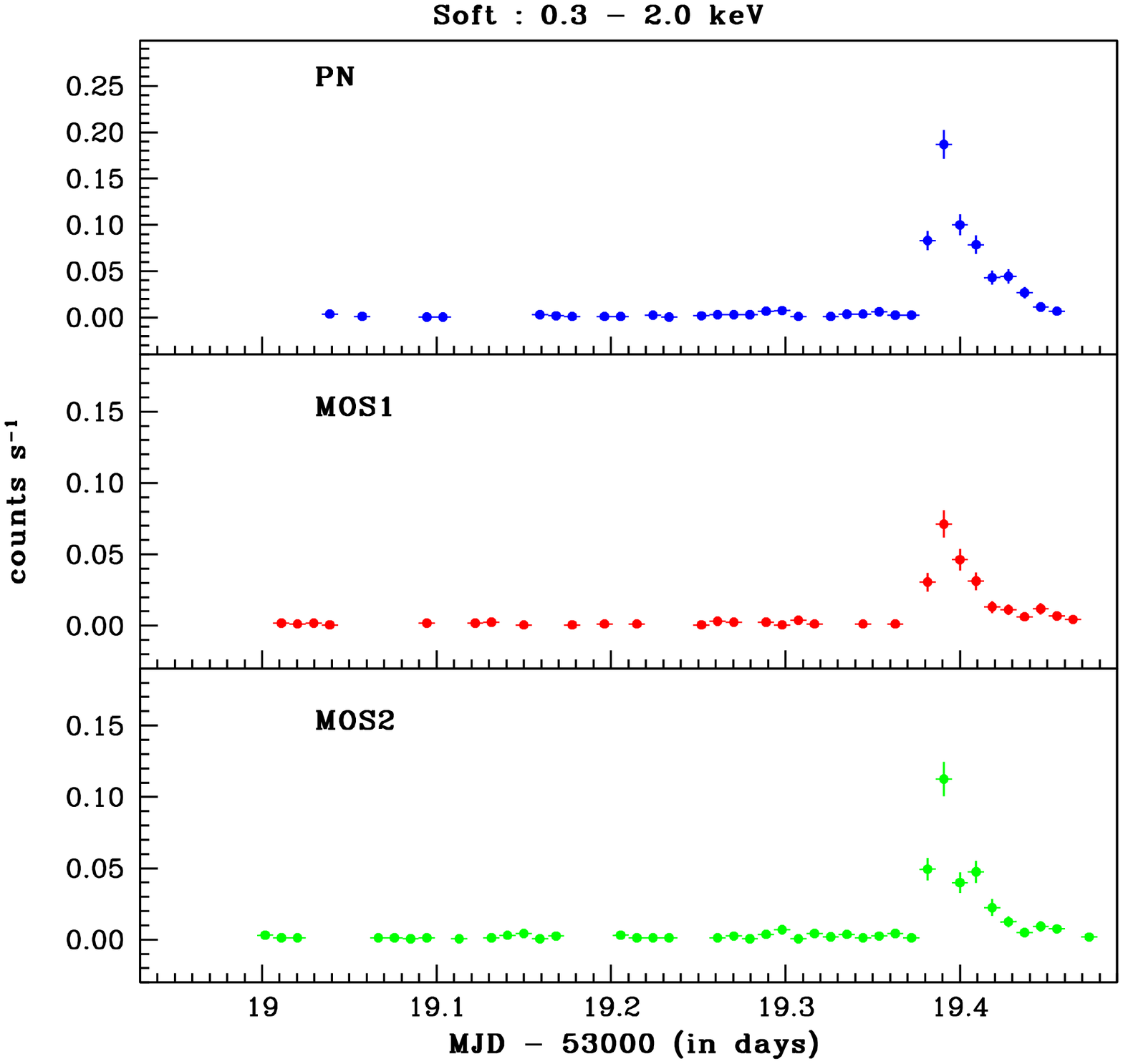}
\includegraphics[width=85mm]{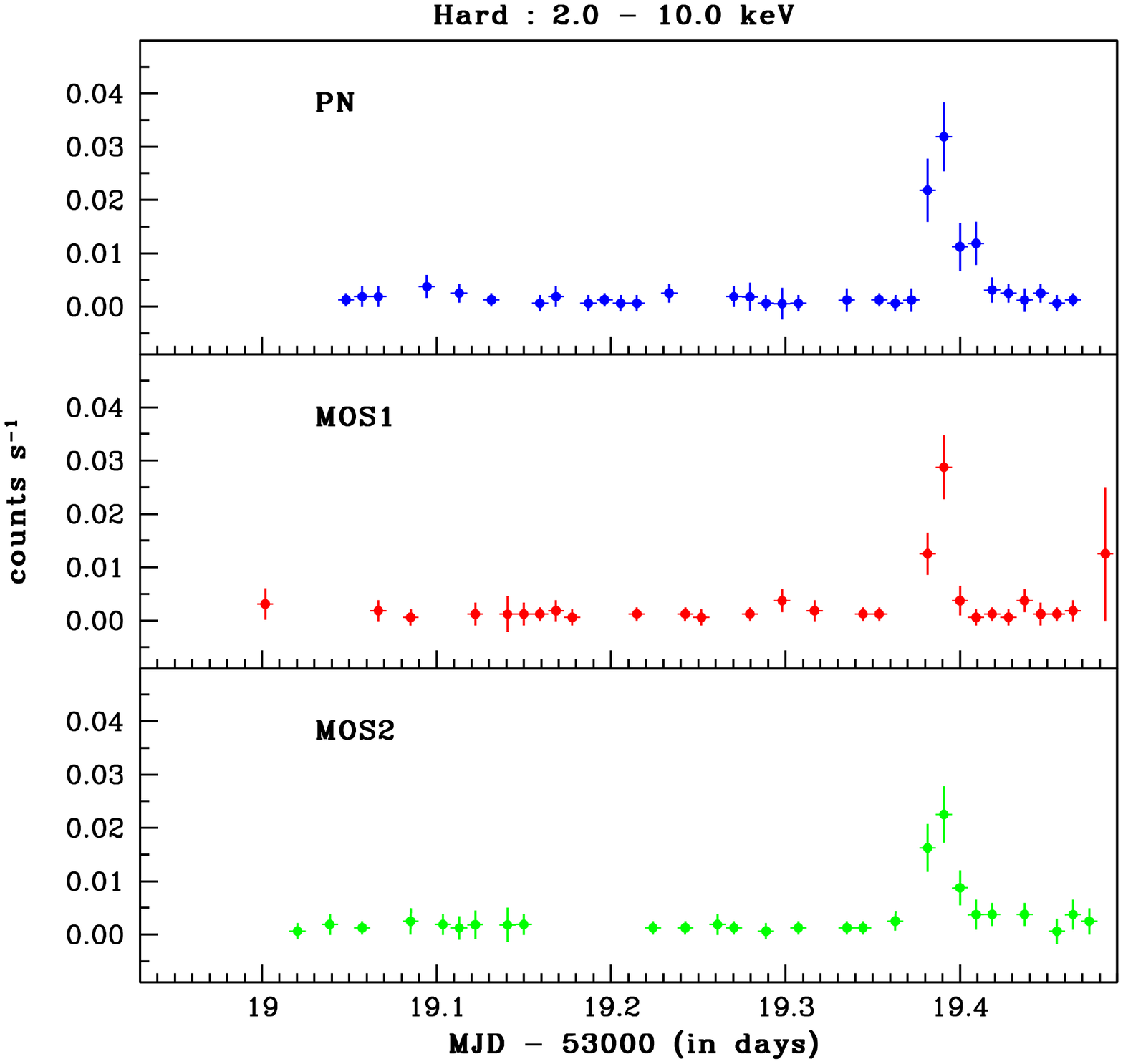}
\caption{Background corrected lightcurves in EPIC detectors in energy bands $\it {left}$ : soft band-- S (0.3 - 2.0 keV) and $\it {right}$ : hard band--H (2.0 - 10.0 KeV).}
\label{fig:lightcurves_sub}
\end{figure*}

The optical V-band magnitudes of 3XMM J014528.9+610729  for each exposure time are plotted in the lower panel of Figure~\ref{fig:lightcurves_sub},
 where dashed line represents the mean V-band magnitude of  3XMM J014528.9+610729.
The flare  and  quiescent state regions of X-ray flare are shown by dotted lines.
The flux in V-band show small enhancement during the flare, however the enhancement is within 2$\sigma$ significance level due to the large uncertainties in V-band magnitudes.

\subsection{X-ray Spectra}
\label{sec:spec}
\begin{figure*}
\includegraphics[width=58mm,angle=270]{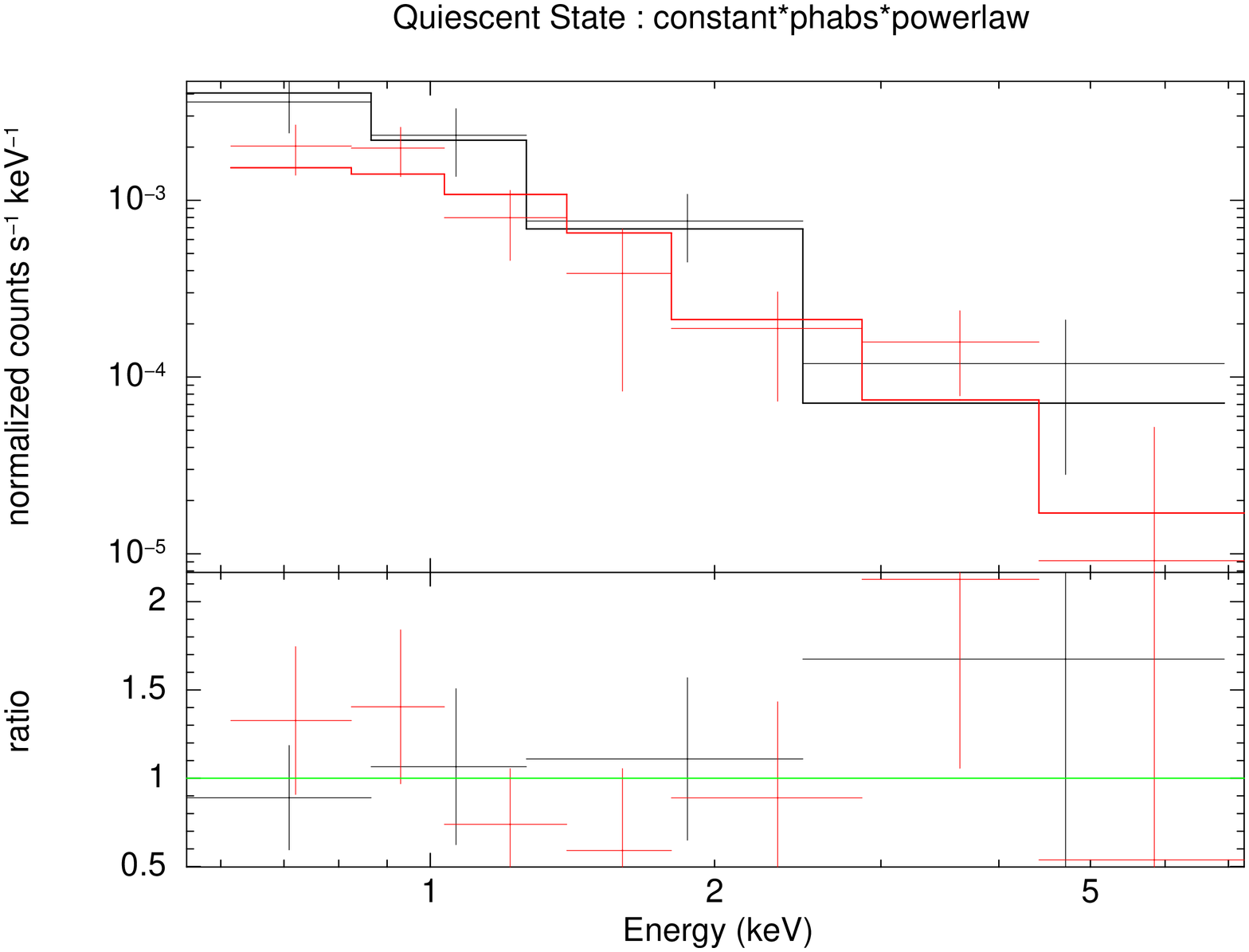}
\vspace{1cm}
\includegraphics[width=58mm,angle=270]{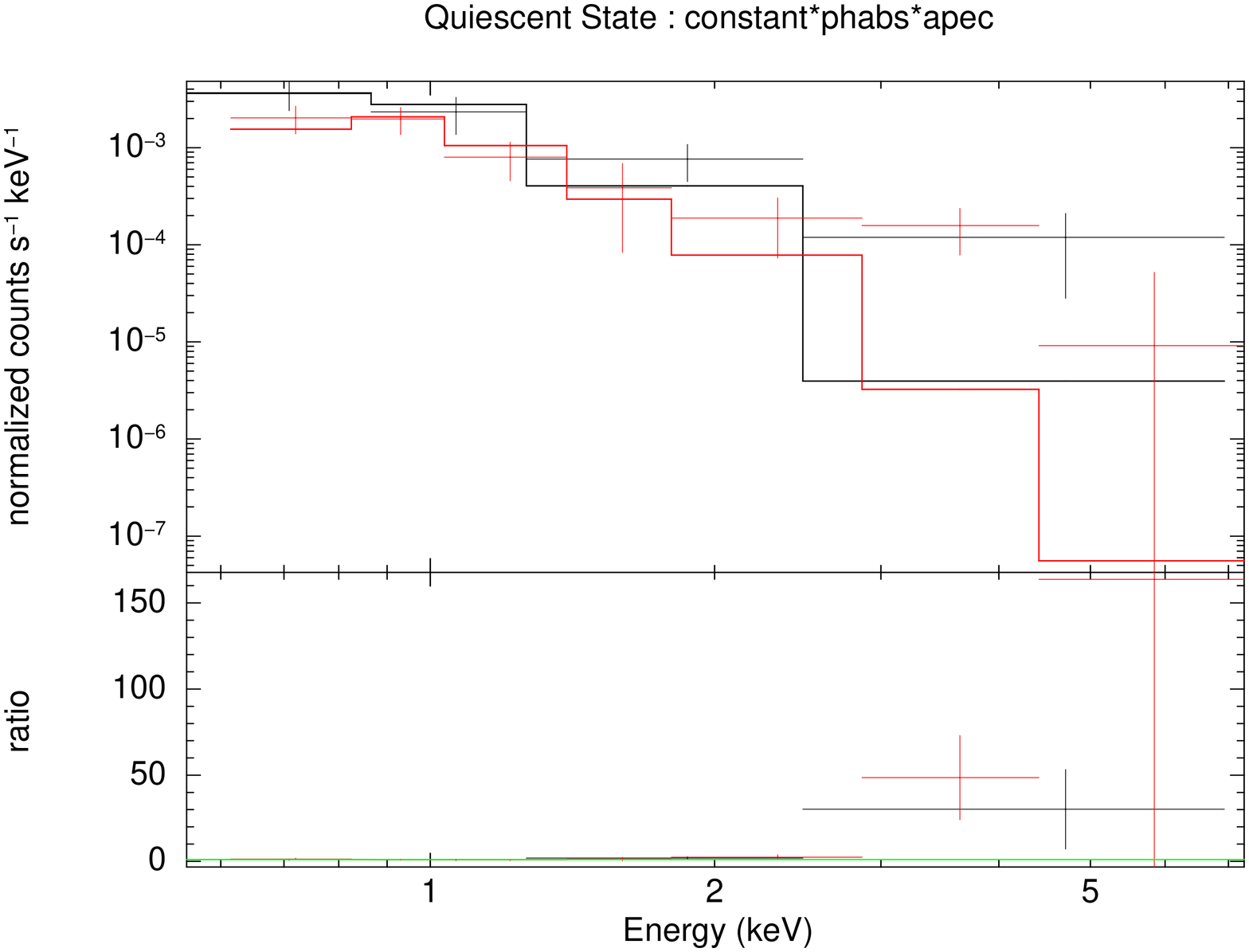}
\vspace{1cm}
\includegraphics[width=58mm,angle=270]{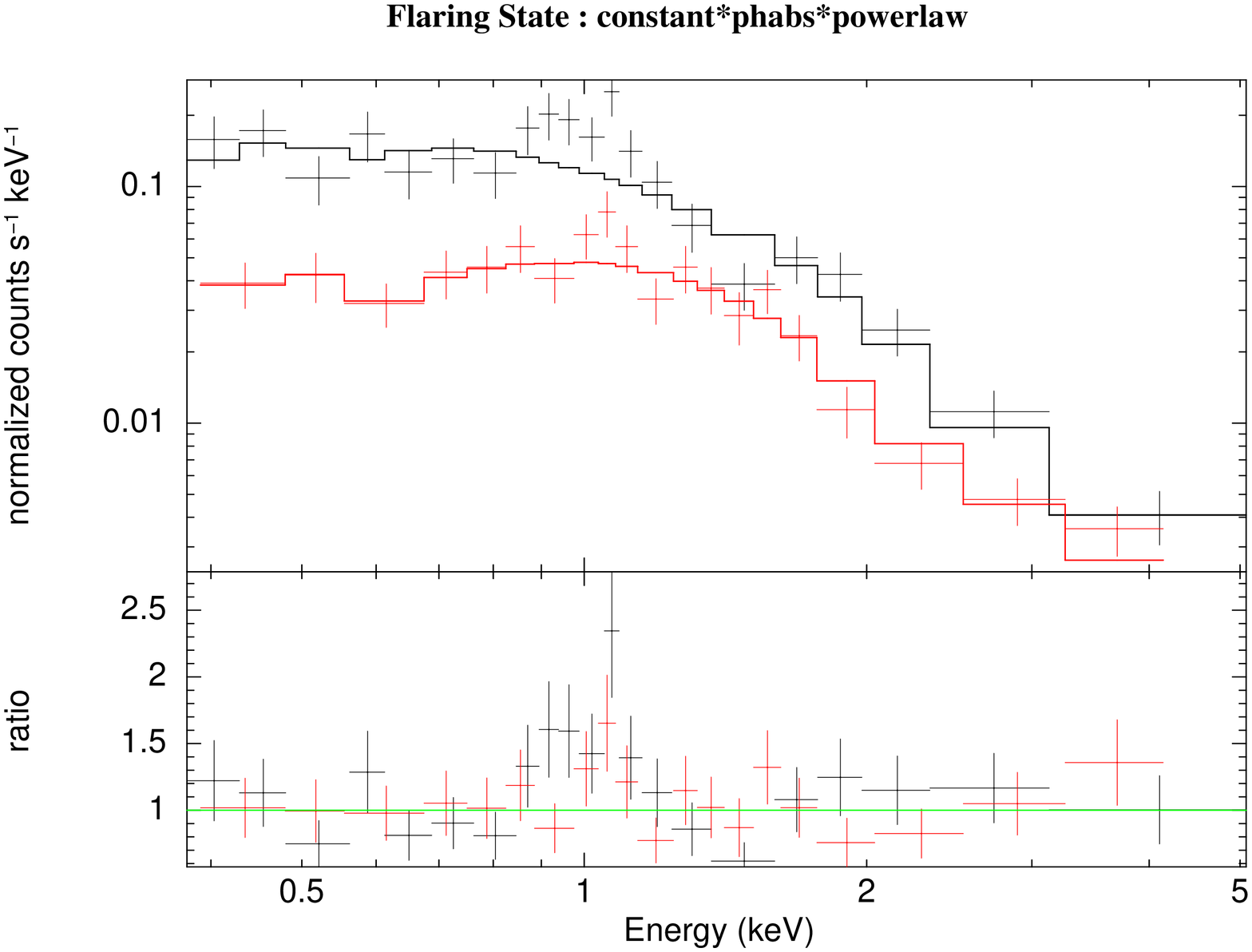}
\includegraphics[width=58mm,angle=270]{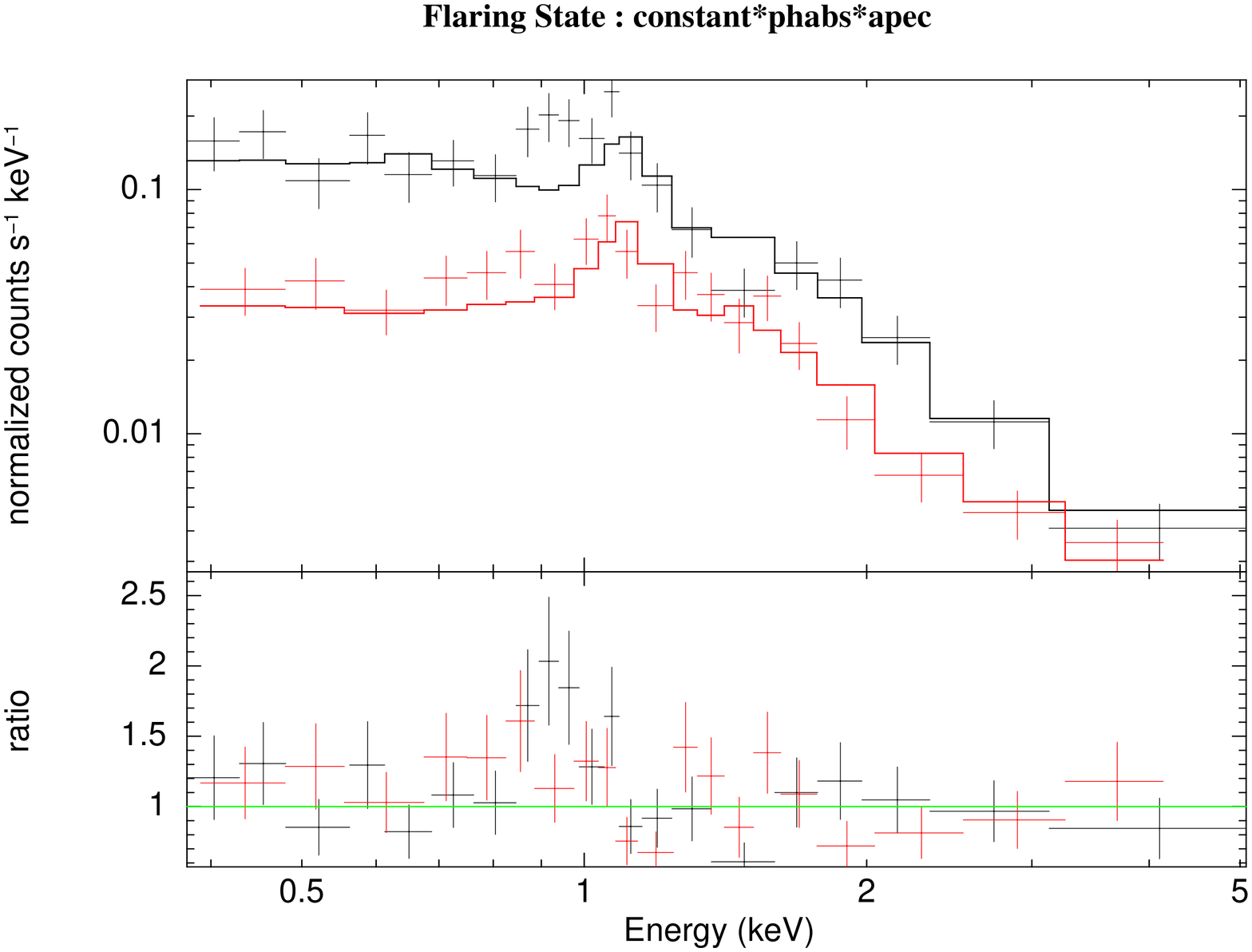}
\includegraphics[width=58mm,angle=270]{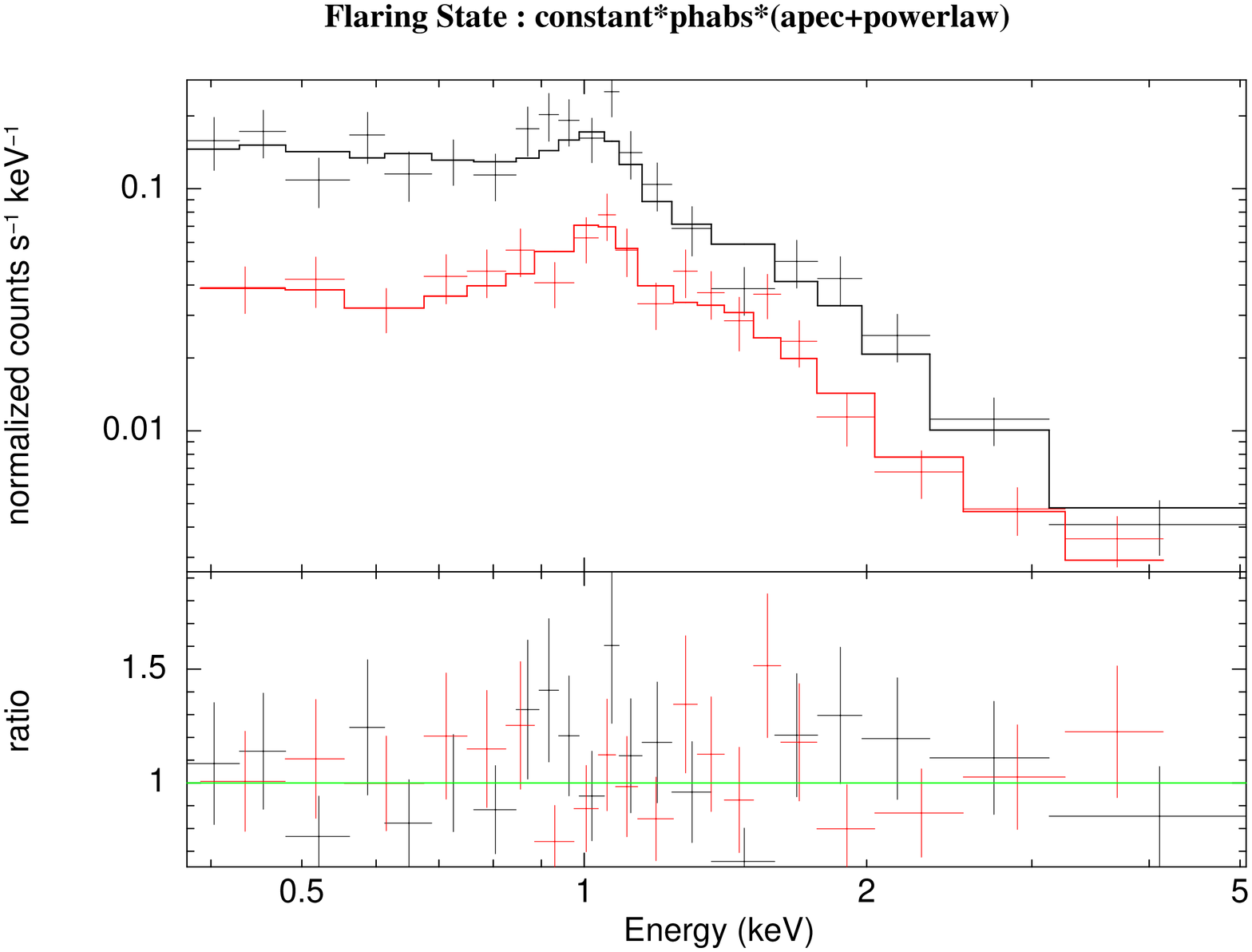}
\vspace{1cm}
\caption{X-ray spectra in PN detector (black colour; available only in electronic form) and combined (MOS1+MOS2) spectra in MOS detector (red colour; available only in electronic form)  with best-fitted models for quiescent and flaring states. The $\chi^2$ distribution in terms of ratio are given in lower sub panels.}
\label{fig:spec} 
\end{figure*}

The X-ray spectra for the flaring state "Fl" and the quiescent state "Q" were generated independently in the total energy band.
The photon redistribution as well as the ancillary matrices were computed using the SAS task {\sc rmfgen} and {\sc arfgen}. 
The data from MOS1 and MOS2 CCDs were combined using {\sc Heasoft}\footnote{http://heasarc.gsfc.nasa.gov/lheasoft} task {\sc addspec}.
The PN and combined MOS spectra were rebinned to have at least 20 counts per bin.
The spectra in  the flare state and  the quiescent state were fitted with three different models --
(a) non-thermal, (b) thermal and (c) combination of thermal and non-thermal models.
The thermal model Astrophysical Plasma Emission Code ({\sc{Apec}}) version 1.10 modeled by \citet{sim+01} and 
the non-thermal  {\sc Power-law} model  are used for global fitting of X-ray spectra.  
The Galactic photoelectric absorption of X-rays was accounted for by using a 
multiplicative model {\sc phabs} in {\sc Xspec} \citep{bal+92}.
The spectra in PN and MOS detectors were fitted  simultaneously.
The relative cross-calibration of PN and MOS detectors was taken care by introducing 
a floating normalization constant in the model during the fitting process.
The best fit parameters of {\sc Power-law} model, {\sc Apec} model and combined ({\sc Apec + Power-law}) model 
are derived using $\chi^2$-minimization technique in {\sc Xspec} version 12.8.0 and given in Table~\ref{tab:specfit}.
The X-ray spectra with best-fitted models are shown in Figure~\ref{fig:spec} for the quiescent and the flaring states.
Due to poor count statistics, we could not fit the  Combined ({\sc Apec + Power-law}) model during the quiescent state.

The hydrogen column density, $\rm{N_H}$, along the line of sight of the X-ray source is estimated to be $\rm {6.7\pm1.2~\times10^{21}~cm^{-2}}$ using 
{\sc Heasoft tool\footnote{http://heasarc.gsfc.nasa.gov/cgi-bin/Tools/w3nh/}} \citep[LAB map;][]{kal+05} with cone radius of 1 degree.
We can not constrain other parameters in spectral fitting 
while freezing the value of $\rm{N_H}$ with $\rm{6.7\pm1.2~\times10^{21}~cm^{-2}}$  in quiescent as well as flaring states.
Therefore, we used $\rm{N_H}$ as a free parameter during spectral fitting.
Using {\sc Power-law} model, the best fitted values of $\rm{N_H}$ were found to be $\rm{>5.0 \times10^{21}~cm^{-2}}$ 
and $\rm{1.4^{+0.4}_{-0.3}\times10^{21}~cm^{-2}}$
for quiescent and flaring states, respectively.
Using {\sc Apec} model, the best fit values of $\rm{N_H}$ were found to be  $\rm{4.8^{+1.5}_{-1.8}\times10^{21}~cm^{-2}}$ 
and  $\rm{>5.0 \times10^{21}~cm^{-2}}$ for the quiescent and the flaring states, respectively.
These values of  $\rm{N_H}$ are nearly similar to that of estimated with LAB map within $\rm {2\sigma}$ limits,
however, it is lower during the flaring state with {\sc Power-law} model.

The value of $\rm{kT}$ were derived to be $\rm{0.54^{+0.13}_{-0.17}}$ keV and $\rm{3.18^{+0.31}_{-0.29}}$ keV
during quiescent and flaring states, respectively.
The best fit values of power-law indices were estimated to be $\rm{2.27^{+0.45}_{-0.37}}$ and $\rm{2.49^{+0.19}_{-0.18}}$
for the quiescent and the flaring states, respectively.  
As the possible mechanism of the flare is not known, we have also fitted the flare spectrum by {\sc Apec+Powerlaw} model. 
This gave a relatively lower temperature ($\rm{1.33^{+0.20}_{-0.09}}$ keV) of the thermal plasma and harder power-law index 
($\rm{1.93^{+0.24}_{-0.23}}$) as compared to what was obtained with the {\sc Power-law} model only (see Table~\ref{tab:specfit}). 
The $\rm{\chi^2_\nu}$ was improved significantly while fitting the spectra with the combined ({\sc Apec+Powerlaw}) model.

\begin{table*}
\normalsize
\caption{The best fit parameters of thermal {\sc Apec} and {\sc Power-law} model.} 
\begin{tabular}{lrrr}
\hline
State         & Quiescent (Q)                               &             & Flare (Fl) \\
\hline
Model                                  & \multicolumn{3}{c}{constant*phabs*powerlaw} \\
\hline
$\rm{N_H}$  ($\rm{10^{22}}~cm^{-2}$)                  &  $>0.50$                      &  &  $0.14^{+0.04}_{-0.03}$               \\ 
Power-law                                             &  $2.27^{+0.45}_{-0.37}$       &  &  $2.49^{+0.19}_{-0.18}$               \\
Normalization ($\rm{10^{-5}}$)                        &  $0.37^{+0.08}_{-0.08}$       &  &  $25.85^{+3.96}_{-3.39}$               \\
Constant factor                                       &  $2.02^{+0.77}_{-0.53}$       &  &  $1.47^{+0.11}_{-0.10}$               \\ 
Log(Flux)  ($\rm{erg~s^{-1}}~cm^{-2}$) [0.3-10.0 keV] & $-13.73^{+0.06}_{-0.06}$   &  &  $-11.90^{+0.01}_{-0.01}$          \\ 
Log(Flux)  ($\rm{erg~s^{-1}}~cm^{-2}$) [0.3-2.0 keV]  & $-13.91^{+0.16}_{-0.25}$   &  &  $-12.03^{+0.01}_{-0.02}$           \\
Log(Flux)  ($\rm{erg~s^{-1}}~cm^{-2}$) [2.0-5.0 keV]  & $-14.40^{+0.16}_{-0.25}$   &  &  $-12.66^{+0.01}_{-0.02}$           \\
Log(Flux)  ($\rm{erg~s^{-1}}~cm^{-2}$) [5.0-10.0 keV] & $-14.62^{+0.16}_{-0.25}$   &  &  $-12.96^{+0.01}_{-0.02}$           \\
$\chi^2_\nu$ (dof)                          &  0.66 (7)                  & &    1.30 (37)                      \\
\hline
Model                                  & \multicolumn{3}{c}{constant*phabs*apec} \\
\hline
$\rm{N_H}$  ($\rm{10^{22}}~cm^{-2}$)                  & $0.48^{+0.15}_{-0.18}$        & & $>0.5$                             \\  
kT (keV)                                              & $0.54^{+0.13}_{-0.17}$        & & $3.18^{+0.31}_{-0.29}$             \\
Normalization ($\rm{10^{-5}}$)                        & $1.26^{+1.66}_{-0.67}$        & & $47.19^{+2.57}_{-2.56}$             \\
Constant factor                                       & $1.87^{+0.74}_{-0.50}$        & & $1.39^{+0.10}_{-0.10}$             \\ 
Log(Flux)  ($\rm{erg~s^{-1}}~cm^{-2}$) [0.3-10.0 keV] & $-13.40^{+0.06}_{-0.07}$   & &  $-12.09^{+0.01}_{-0.01}$       \\   
Log(Flux)  ($\rm{erg~s^{-1}}~cm^{-2}$) [0.3-2.0 keV]  & $-13.40^{+0.06}_{-0.07}$   & &  $-12.37^{+0.01}_{-0.02}$        \\  
Log(Flux)  ($\rm{erg~s^{-1}}~cm^{-2}$) [2.0-5.0 keV]  & $-15.29^{+0.07}_{-0.06}$   & &  $-12.60^{+0.01}_{-0.02}$        \\  
Log(Flux)  ($\rm{erg~s^{-1}}~cm^{-2}$) [5.0-10.0 keV] & $-17.81^{+0.06}_{-0.07}$   & &  $-12.86^{+0.01}_{-0.02}$        \\  
$\chi^2_\nu$ (dof)                          &  1.30 (7)                  & &   1.54 (37)                       \\
\hline
Model                                  & \multicolumn{3}{c}{constant*phabs*(apec+powerlaw)} \\
\hline
$\rm{N_H}$  ($\rm{10^{22}}~cm^{-2}$)                  &                          & &  $<0.08$                             \\  
kT (keV)                                              &                          & &  $1.33^{+0.20}_{-0.09}$             \\
Normalization  (thermal;$\rm{10^{-5}}$)               &                          & &  $7.71^{+3.35}_{-2.05}$             \\
Power-law                                             &                          & &  $1.93^{+0.24}_{-0.23}$               \\
Normalization (powerlaw;$\rm{10^{-5}}$)               &                          & &  $13.17^{+3.34}_{-2.71}$               \\
Constant factor                                       &                          & &  $1.43^{+0.11}_{-0.10}$             \\ 
Log(Flux)  ($\rm{erg~s^{-1}}~cm^{-2}$) [0.3-10.0 keV] &                          & &  $-12.03^{+0.02}_{-0.02}$       \\   
Log(Flux)  ($\rm{erg~s^{-1}}~cm^{-2}$) [0.3-2.0 keV]  &                          & &  $-12.28^{+0.01}_{-0.02}$        \\  
Log(Flux)  ($\rm{erg~s^{-1}}~cm^{-2}$) [2.0-5.0 keV]  &                          & &  $-12.63^{+0.02}_{-0.02}$        \\  
Log(Flux)  ($\rm{erg~s^{-1}}~cm^{-2}$) [5.0-10.0 keV] &                          & &  $-12.77^{+0.03}_{-0.03}$        \\  
$\chi^2_\nu$ (dof)                          &                          & &   0.93 (35)                       \\
\hline
\end{tabular}
\label{tab:specfit}
\end{table*} 

\begin{figure}
\subfigure[]{\includegraphics[width=85mm]{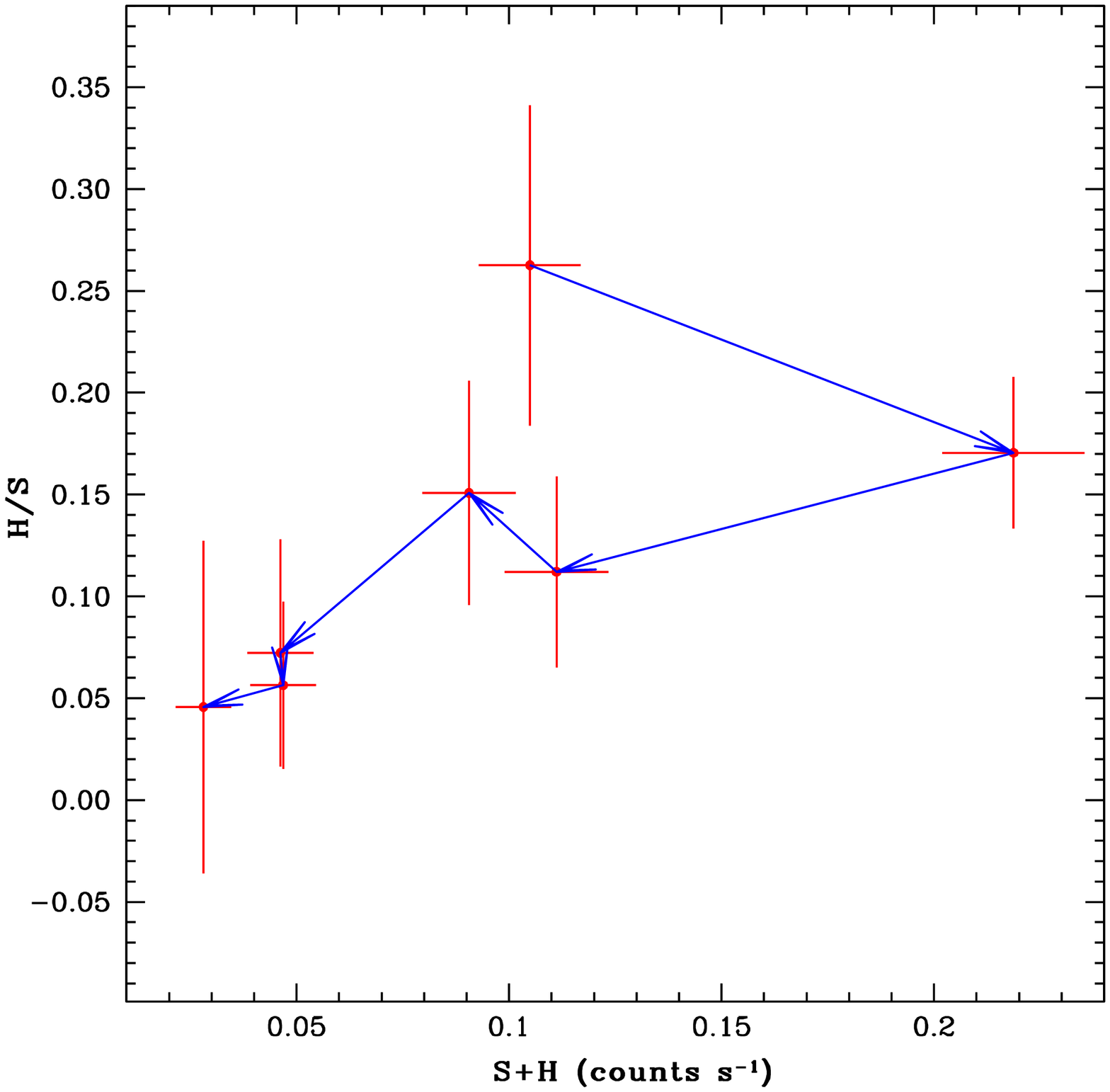}}
\subfigure[]{\includegraphics[width=85mm]{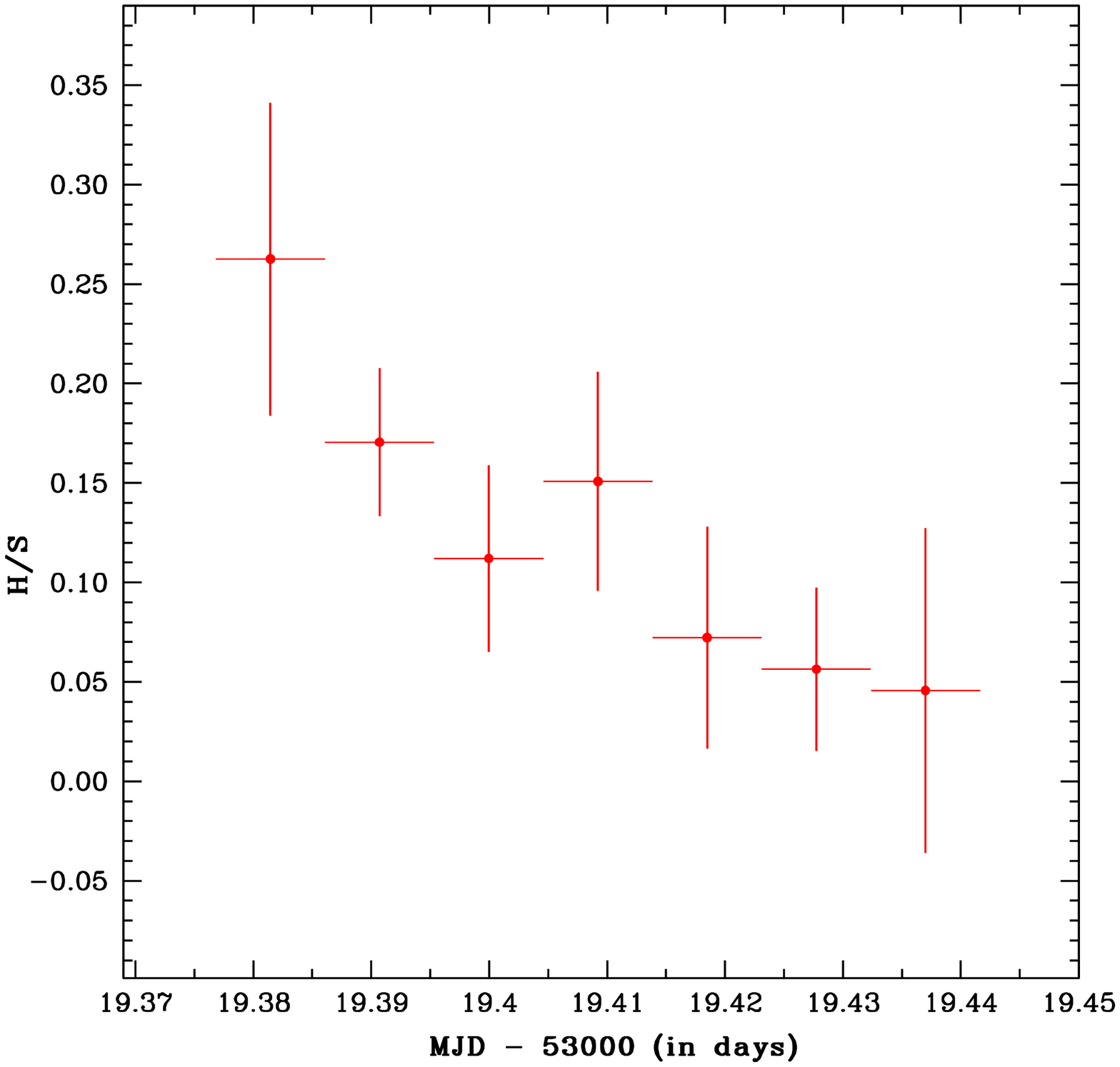}}
\caption{ (a) Relation between hardness ratio (H/S) and total count rates (H+S) during the flaring state. The sense of flux evolution is shown by the arrows. (b) Evolution of hardness ratio (H/S)  during the flaring state.  }
\label{fig:evolution}
\end{figure}
 
\section{Discussion}\label{sec:discuss}
An X-ray flare has been detected from 3XMM J014528.9+610729 during the observations of star cluster NGC 663 by {\sc XMM-Newton}.
The colour-colour diagrams are powerful diagnostic tools in such cases where no spectroscopic information about of the sources is known.
The location of the X-ray source 3XMM J014528.9+610729  in the colour-colour plane is used to identify the X-ray source in
optical and infrared bands. On the basis of the optical colour-colour diagram, the possibility of the source for being a QSO is ruled out.
Comparing the near and mid infrared  properties of the source with the stars and galaxies using 2MASS and WISE data, 
it has been evident that the source is a candidate normal spiral galaxy.
It makes the X-ray source  3XMM J014528.9+610729 very interesting to study as the flaring event from the non-active galaxies are very rare.   
The behavior of the X-ray flare and  possible scenarios for the generation of the X-ray flare were further investigated.

As we do not know the exact mechanism and site of radiation emission, so we used phenomenological models to fit the spectra in the quiescent 
and the flaring states. The fitting of the quiescent state spectrum with {\sc Apec} thermal model gave a temperature of $\rm{0.54^{+0.13}_{-0.17}}$ keV
of the emitting plasma, whereas the fitting with {\sc Power-law} model gave a power-law index of $\rm{2.27^{+0.45}_{-0.37}}$.
The value of equivalent hydrogen column density is consistent in both the cases.
But the poor statistics of data in the quiescent state makes it difficult to distinguish the two models.
Therefore, any further constrain can not be imposed on the quiescent state spectrum.

In case of the flaring state, we fitted the spectrum with {\sc Apec} model and {\sc Power-law} model. 
The {\sc Apec} model gave a temperature of $\rm{3.18^{+0.31}_{-0.29}}$ keV implying substantial heating of the plasma during the flaring 
process. The {\sc Power-law} model gave a spectral index of $\rm{2.49^{+0.19}_{-0.18}}$.
This implies that if the basic origin of radiation in the source is entirely due to some non thermal process then the spectral shape
of the time averaged flare spectrum does not change much. The time averaged flare spectrum does not show significant steepening in the spectrum.
The fitting of the flaring state spectrum improves when it is fitted with {\sc Apec+Power-law} model. In such a condition 
the temperature of the plasma was obtained to be $\rm{1.33^{+0.20}_{-0.09}}$ keV, which is cooler than that of obtained from {\sc Apec} model only.
The power-law index is found to be $\rm{1.93^{+0.24}_{-0.23}}$, which
is harder than that of obtained in the pure {\sc Power-law} case. Thus if it so happens that during the flare
the plasma is heated and a fraction of the thermal particles of the plasma are accelerated to higher energies by some
acceleration process generating a harder spectrum, then the radiation emission can be due to some hybrid of thermal and non-thermal distribution of
particles.

To have a better understanding of the flux and spectral evolution during the flare, we studied the time variation 
of the hardness ratio (H/S) and the hardness ratio-flux correlation. The time variation of the hardness ratio shows that the hardness
reduces as the flare progresses, implying the softening of the spectrum. The hardness--flux correlation reveals a clockwise sense as the hardness 
evolves with the flux. Such sense actually implies that the lower energy radiation lags the higher energy emission 
during the radiation emission process. This kind of phenomena is observed in the case of optically thin emission from blazar jets 
\citep[e.g.,][]{kir+98, fos+00a, fos+00b, bhatta+05, zhang+06}.
Here, also if the radiation emission takes place in an optically thin emission region then the clockwise sense  of 
hardness count correlation can be explained if the high energy particles are injected in the emission 
region within a very short time scale, and then the particles are allowed to cool by emitting radiation.

The decay appears nearly exponential, therefore, it can not be associated with  
the flares observed due to the tidal disruption of a star by a SMBHs in the nuclei of galaxies because 
in such cases the flux decays with time following a power-law \citep[$\rm{t^{-5/3}}$; e.g.,][]{rees88,gre+00,esq+07,cap+09,sax+12}.
The thermal components which are found in the present analysis are harder compared 
to the component detected in the case of the flares due to the 
tidal disruption of a star by a SMBHs in the 
nuclei of galaxies, which occurs in the temperature range  0.04-0.1 keV \citep[for NGC 5905;][]{kom+02}. 
However, the possibility of the X-ray source 3XMM J014528.9+610729 for being a Galactic foreground object could not be completely ruled out 
due to the lack of optical spectroscopic data.

\section{Conclusions}
\label{sec:con}
We have detected an X-ray flare from the  X-ray source 3XMM J014528.9+610729 which is  serendipitously observed
during the X-ray observations of the open star cluster NGC 663 from {\sc XMM-Newton}.
The identification of the X-ray source using multiwavelength data sets in optical and infrared bands 
has been performed, and the spectral and temporal characteristics of the source during the quiescent and the flaring states have been investigated.
The main conclusions of present analysis are as follows.

\begin{itemize}
\item{The X-ray source is found to be a candidate spiral galaxy using colour-colour information 
in optical, and near and mid infrared bands.}
\item{The flare has highly asymmetrical time structure with the FRED shape, and the rise and decay 
times of the flare are estimated to be  $\sim$1.6 ks and $\sim$4.0 ks, respectively. }
\item{The spectrum of the source during the  quiescent state is fitted with thermal {\sc Apec} model and also
with non-thermal {\sc Power-law} model. Due to the poor statistics of the data in the quiescent state 
no firm conclusion can be drawn regarding the nature of the source.}
\item{In the flaring state, the spectrum can be best fitted with a spectral model combining two models ({\sc Apec+Powerlaw }). }
\item{The variation of hardness with flux indicates a clockwise structure which implies a soft lag in the emission process.}
\item{As the specific nature of the flare mechanism is not known for a spiral galaxy, so regular monitoring of the
source 3XMM J014528.9+610729 is required in X-rays and other wavebands to have an 
improved understanding of the source emission mechanism and link between active and normal galaxies. }
\end{itemize}

\section*{acknowledgments}
Authors are thankful to Roberto J. Assef and Roser Pello for their help to use their software for fitting SED.
This publication makes use of data from the Two Micron All Sky Survey, which is a joint project of
the University of Massachusetts and the Infrared Processing and Analysis Center/California
Institute of Technology, funded by the National Aeronautics and Space Administration and
the National Science Foundation,  data products from the Wide-field Infrared Survey Explorer, 
which is a joint project of the University of California, Los Angeles, and the Jet Propulsion Laboratory/California Institute of Technology, funded by the National Aeronautics and Space Administration
and SDSS which is by the Alfred P. Sloan Foundation, the Participating Institutions, the National Science Foundation, the U.S. Department of Energy, 
the National Aeronautics and Space Administration, the Japanese Monbukagakusho, the Max Planck Society, and the Higher Education Funding Council for England.
Data obtained from the High Energy Astrophysics Science Archive Research Center (HEASARC), provided by NASA’s Goddard
The Space Flight Center has also been used in the present study.
HB is thankful for the financial support for this work through the INSPIRE faculty fellowship granted by the Department of Science \& Technology  India, and
R. Koul for his support to work and to pursue DST-INSPIRE position at
ApSD, BARC, Mumbai. 

\bibliographystyle{mn2e}
\bibliography{ms}
\clearpage
\input{sub.tex}

\end{document}

%% file: def.tex
\def\gsim{\;\rlap{\lower 2.5pt\hbox{$\sim$}}\raise 1.5pt\hbox{$>$}\;}
\def\lsim{\;\rlap{\lower 2.5pt\hbox{$\sim$}}\raise 1.5pt\hbox{$<$}\;}
\def\la{\mathrel{\hbox{\rlap{\hbox{\lower4pt\hbox{$\sim$}}}\hbox{$<$}}}}
\def\ga{\mathrel{\hbox{\rlap{\hbox{\lower4pt\hbox{$\sim$}}}\hbox{$>$}}}}

\def\arcmin{\hbox{$^\prime$}}
\def\arcsec{\hbox{$^{\prime\prime}$}}

\def\micron{\hbox{$\mu$m}}

%% file: sub.tex
%\documentclass{article}
%\usepackage{graphicx}
%\usepackage{subcaption}

%\begin{document}
\renewcommand{\thefigure}{\Alph{figure}}
\setcounter{figure}{0}

\begin{figure*}
\includegraphics[width=150mm]{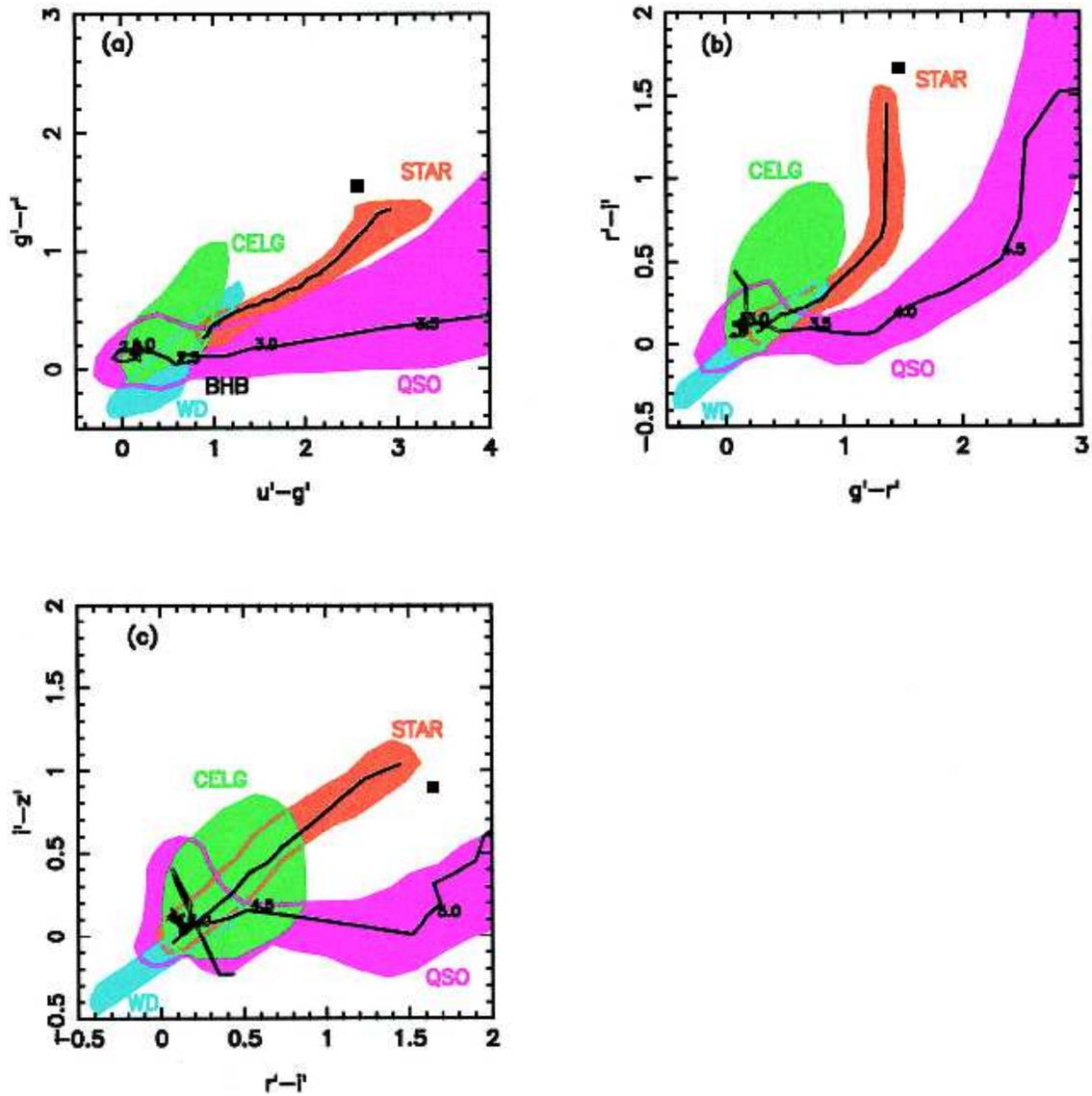}
\caption{Color-color diagrams from  Fan (1999), 
(a) ($u^\prime$ - $g^\prime$) vs  ($g^\prime$ - $r^\prime$), 
(b) ($g^\prime$ - $r^\prime$) vs  ($r^\prime$ - $i^\prime$) and 
(c) ($r^\prime$ - $i^\prime$) vs  ($r^\prime$ - $z^\prime$). 
The locations of objects with different classes:
normal Galactic stars (STAR), white dwarfs (WD), quasars (QSO), 
compact emission-line galaxies (CELGs) and halo blue horizontal branch
stars (BHBs) are shown. 
The solid lines going through the stellar regions are the best-fitted stellar locus points using the algorithm of Newberg \& Yanny (1997). 
The lines going through the quasar regions are
the median quasar tracks as a function of redshift.
The location of 3XMM J014528.9+610729 
in this color-color plane is shown by a symbol of square. 
}
\end{figure*}

\begin{figure*}
\includegraphics[width=140mm,height=120mm]{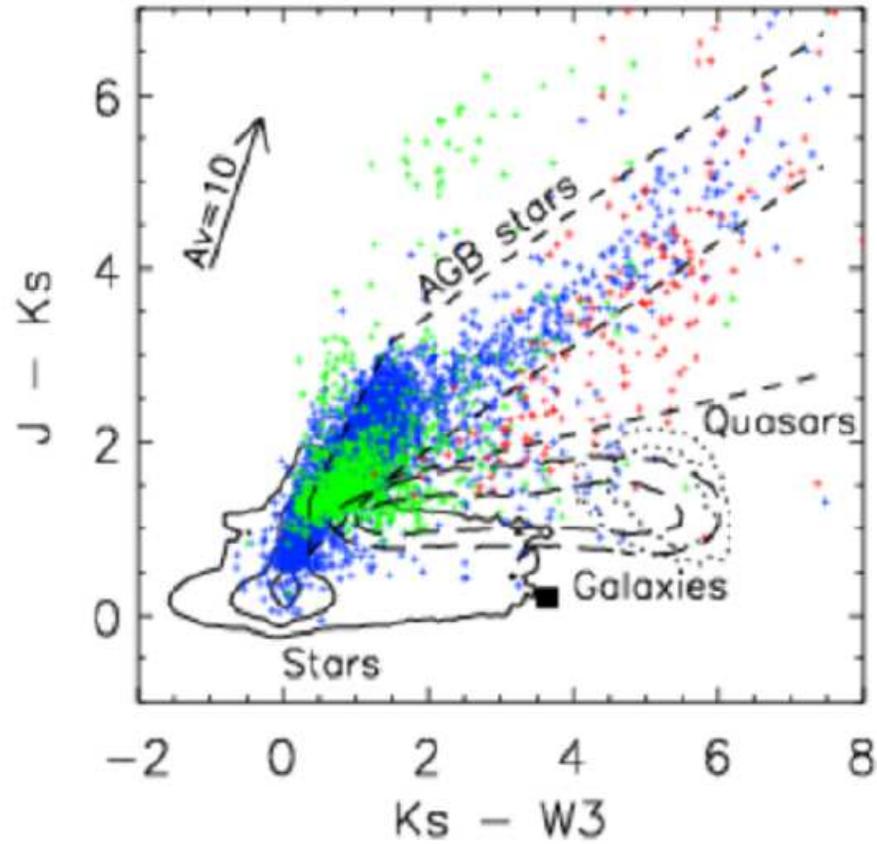}
\caption{Color-color diagram, ($K_s$ -W3) vs (J-$K_s$) from Tu \& Wang (2013). 
Plus symbols : blue represents C-rich AGB stars, green represents AGB stars and red  represents O-rich stars. 
The solid contours show 1-, 2-, and 3-$\sigma$ regions
of the normal stars, long-dashed contours 1- and 2-$\sigma$ regions of the galaxies, and dotted contours 
1- and 2-$\sigma$ regions of the quasars. The dashed lines define the region for the AGB stars.
The direction of extinction is shown by the arrow.  
The symbol of square represents the location of 3XMM J014528.9+610729  in color-color plane. }
\end{figure*}

\begin{figure*}
\includegraphics[width=130mm,height=120mm]{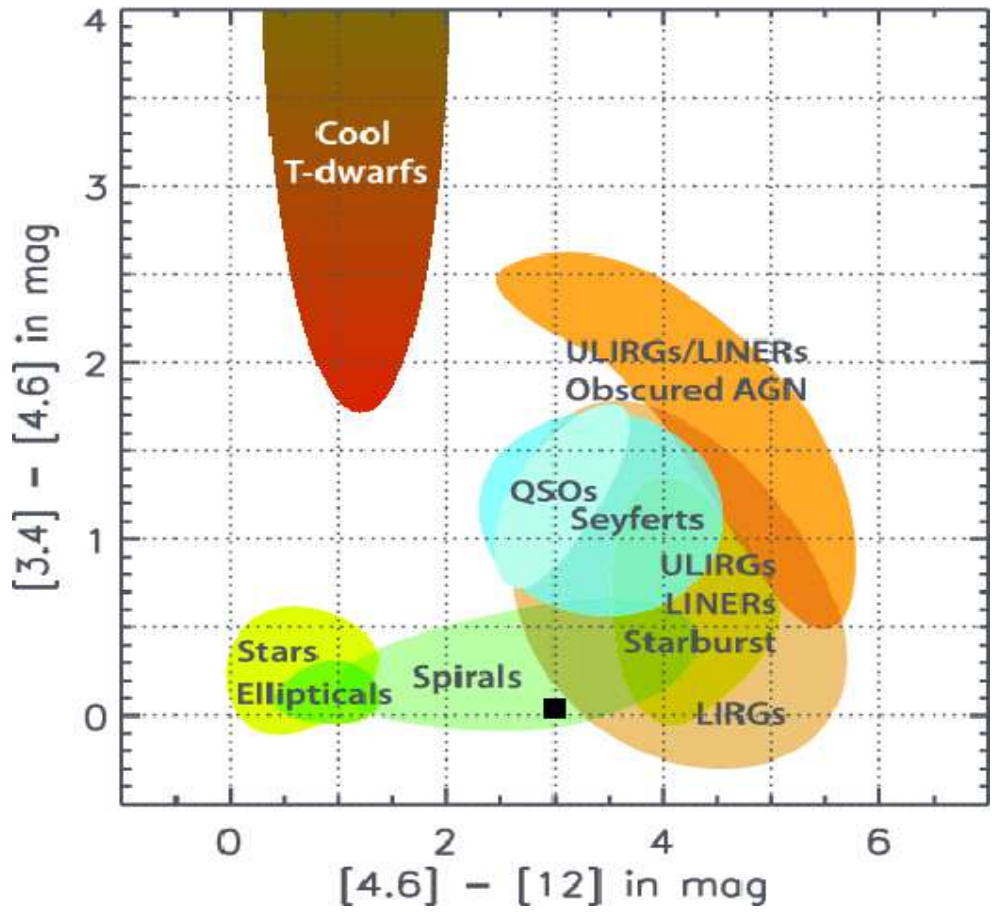}
\caption{Color-color diagram (W2-W3) vs (W1-W2) : figure
is taken from Wright et al. (2010). , Different regions occupied in color-color plane by stars, brown dwarfs, different type of
galaxies and AGNs. of 2MASS and WISE bands showing the locations of objects with different classes. The symbol of square
represents the location of 3XMM J014528.9+610729 in color-color plane.}
\end{figure*}

%\end{document}

%% file: ms.bbl
\begin{thebibliography}{45}
\expandafter\ifx\csname natexlab\endcsname\relax\def\natexlab#1{#1}\fi

\bibitem[{{Abazajian} {et~al}\mbox{.}(2009){Abazajian}, {Adelman-McCarthy},
  {Ag{\"u}eros}, {Allam}, {Allende Prieto}, {An}, {Anderson}, {Anderson},
  {Annis}, {Bahcall}, \& et~al.}]{aba+09}
{Abazajian} K.~N. {et~al.}, 2009, \apjs, 182, 543

\bibitem[{{Assef} {et~al}\mbox{.}(2008){Assef}, {Kochanek}, {Brodwin}, {Brown},
  {Caldwell}, {Cool}, {Eisenhardt}, {Eisenstein}, {Gonzalez}, {Jannuzi},
  {Jones}, {McKenzie}, {Murray}, \& {Stern}}]{ass+08}
{Assef} R.~J. {et~al.}, 2008, \apj, 676, 286

\bibitem[{{Assef} {et~al}\mbox{.}(2010){Assef}, {Kochanek}, {Brodwin}, {Cool},
  {Forman}, {Gonzalez}, {Hickox}, {Jones}, {Le Floc'h}, {Moustakas}, {Murray},
  \& {Stern}}]{ass+10}
{Assef} R.~J. {et~al.}, 2010, \apj, 713, 970

\bibitem[{{Balucinska-Church} \& {McCammon}(1992)}]{bal+92}
{Balucinska-Church} M., {McCammon} D., 1992, \apj, 400, 699

\bibitem[{{Bhatt} {et~al}\mbox{.}(2013){Bhatt}, {Pandey}, {Singh}, {Sagar}, \&
  {Kumar}}]{bhat+13}
{Bhatt} H., {Pandey} J.~C., {Singh} K.~P., {Sagar} R., {Kumar} B., 2013,
  Journal of Astrophysics and Astronomy, 34, 393

\bibitem[{{Bhatt} {et~al}\mbox{.}(2014){Bhatt}, {Pandey}, {Singh}, {Sagar}, \&
  {Kumar}}]{bhat+14}
{Bhatt} H., {Pandey} J.~C., {Singh} K.~P., {Sagar} R., {Kumar} B., 2014,
  Journal of Astrophysics and Astronomy, 35, 39

\bibitem[{{Bhattacharyya} {et~al}\mbox{.}(2005){Bhattacharyya}, {Sahayanathan},
  \& {Bhatt}}]{bhatta+05}
{Bhattacharyya} S., {Sahayanathan} S., {Bhatt} N., 2005, \na, 11, 17

\bibitem[{{Bolzonella} {et~al}\mbox{.}(2000){Bolzonella}, {Miralles}, \&
  {Pell{\'o}}}]{bol+00}
{Bolzonella} M., {Miralles} J.-M., {Pell{\'o}} R., 2000, \aap, 363, 476

\bibitem[{{Brinkman} {et~al}\mbox{.}(1998){Brinkman}, {Aarts}, {den Boggende},
  {Bootsma}, {Dubbeldam}, {den Herder}, {Kaastra}, {de Korte}, {van Leeuwen},
  {Mewe}, {Paerels}, {de Vries}, {Cottam}, {Decker}, {Kahn}, {Rasmussen},
  {Spodek}, {Branduardi-Raymont}, {Guttridge}, {Thomsen}, {Zehnder}, \&
  {Guedel}}]{brin+98}
{Brinkman} A. {et~al.}, 1998, in Science with XMM

\bibitem[{{Cappelluti} {et~al}\mbox{.}(2009){Cappelluti}, {Ajello}, {Rebusco},
  {Komossa}, {Bongiorno}, {Clemens}, {Salvato}, {Esquej}, {Aldcroft},
  {Greiner}, \& {Quintana}}]{cap+09}
{Cappelluti} N. {et~al.}, 2009, \aap, 495, L9

\bibitem[{{Cutri} \& {et al.}(2012)}]{cut+12}
{Cutri} R.~M., {et al.}, 2012, VizieR Online Data Catalog, 2311, 0

\bibitem[{{Cutri} {et~al}\mbox{.}(2003){Cutri}, {Skrutskie}, {van Dyk},
  {Beichman}, {Carpenter}, {Chester}, {Cambresy}, {Evans}, {Fowler}, {Gizis},
  {Howard}, {Huchra}, {Jarrett}, {Kopan}, {Kirkpatrick}, {Light}, {Marsh},
  {McCallon}, {Schneider}, {Stiening}, {Sykes}, {Weinberg}, {Wheaton},
  {Wheelock}, \& {Zacarias}}]{cut+03}
{Cutri} R.~M. {et~al.}, 2003, VizieR Online Data Catalog, 2246, 0

\bibitem[{{den Herder} {et~al}\mbox{.}(2001){den Herder}, {Brinkman}, {Kahn},
  {Branduardi-Raymont}, {Thomsen}, {Aarts}, {Audard}, {Bixler}, {den Boggende},
  {Cottam}, {Decker}, {Dubbeldam}, {Erd}, {Goulooze}, {G{\"u}del}, {Guttridge},
  {Hailey}, {Janabi}, {Kaastra}, {de Korte}, {van Leeuwen}, {Mauche},
  {McCalden}, {Mewe}, {Naber}, {Paerels}, {Peterson}, {Rasmussen}, {Rees},
  {Sakelliou}, {Sako}, {Spodek}, {Stern}, {Tamura}, {Tandy}, {de Vries},
  {Welch}, \& {Zehnder}}]{her+01}
{den Herder} J.~W. {et~al.}, 2001, \aap, 365, L7

\bibitem[{{Edelson} {et~al}\mbox{.}(2002){Edelson}, {Turner}, {Pounds},
  {Vaughan}, {Markowitz}, {Marshall}, {Dobbie}, \& {Warwick}}]{ede+02}
{Edelson} R., {Turner} T.~J., {Pounds} K., {Vaughan} S., {Markowitz} A.,
  {Marshall} H., {Dobbie} P., {Warwick} R., 2002, \apj, 568, 610

\bibitem[{{Edelson} {et~al}\mbox{.}(1990){Edelson}, {Krolik}, \&
  {Pike}}]{ede+90}
{Edelson} R.~A., {Krolik} J.~H., {Pike} G.~F., 1990, \apj, 359, 86

\bibitem[{{Esquej} {et~al}\mbox{.}(2007){Esquej}, {Saxton}, {Freyberg}, {Read},
  {Altieri}, {Sanchez-Portal}, \& {Hasinger}}]{esq+07}
{Esquej} P., {Saxton} R.~D., {Freyberg} M.~J., {Read} A.~M., {Altieri} B.,
  {Sanchez-Portal} M., {Hasinger} G., 2007, \aap, 462, L49

\bibitem[{{Fabbiano}(2006)}]{fab06}
{Fabbiano} G., 2006, Advances in Space Research, 38, 2937

\bibitem[{{Fan}(1999)}]{fan99}
{Fan} X., 1999, \aj, 117, 2528

\bibitem[{{Fossati} {et~al}\mbox{.}(2000{\natexlab{a}}){Fossati}, {Celotti},
  {Chiaberge}, {Zhang}, {Chiappetti}, {Ghisellini}, {Maraschi}, {Tavecchio},
  {Pian}, \& {Treves}}]{fos+04}
{Fossati} G. {et~al.}, 2000{\natexlab{a}}, \apj, 541, 153

\bibitem[{{Fossati} {et~al}\mbox{.}(2000{\natexlab{b}}){Fossati}, {Celotti},
  {Chiaberge}, {Zhang}, {Chiappetti}, {Ghisellini}, {Maraschi}, {Tavecchio},
  {Pian}, \& {Treves}}]{fos+00a}
{Fossati} G. {et~al.}, 2000{\natexlab{b}}, \apj, 541, 153

\bibitem[{{Fossati} {et~al}\mbox{.}(2000{\natexlab{c}}){Fossati}, {Celotti},
  {Chiaberge}, {Zhang}, {Chiappetti}, {Ghisellini}, {Maraschi}, {Tavecchio},
  {Pian}, \& {Treves}}]{fos+00b}
{Fossati} G. {et~al.}, 2000{\natexlab{c}}, \apj, 541, 166

\bibitem[{{Fukugita} {et~al}\mbox{.}(1996){Fukugita}, {Ichikawa}, {Gunn},
  {Doi}, {Shimasaku}, \& {Schneider}}]{fuk+96}
{Fukugita} M., {Ichikawa} T., {Gunn} J.~E., {Doi} M., {Shimasaku} K.,
  {Schneider} D.~P., 1996, \aj, 111, 1748

\bibitem[{{Gabriel} {et~al}\mbox{.}(2004){Gabriel}, {Denby}, {Fyfe}, {Hoar},
  {Ibarra}, {Ojero}, {Osborne}, {Saxton}, {Lammers}, \& {Vacanti}}]{gib+04}
{Gabriel} C. {et~al.}, 2004, in Astronomical Society of the Pacific Conference
  Series, Vol. 314, Astronomical Data Analysis Software and Systems (ADASS)
  XIII, {Ochsenbein} F., {Allen} M.~G., {Egret} D., eds., p. 759

\bibitem[{{Gandhi} {et~al}\mbox{.}(2011){Gandhi}, {Blain}, {Russell},
  {Casella}, {Malzac}, {Corbel}, {D'Avanzo}, {Lewis}, {Markoff}, {Cadolle Bel},
  {Goldoni}, {Wachter}, {Khangulyan}, \& {Mainzer}}]{gan+11}
{Gandhi} P. {et~al.}, 2011, \apjl, 740, L13

\bibitem[{{Greiner} {et~al}\mbox{.}(2000){Greiner}, {Schwarz}, {Zharikov}, \&
  {Orio}}]{gre+00}
{Greiner} J., {Schwarz} R., {Zharikov} S., {Orio} M., 2000, \aap, 362, L25

\bibitem[{{Gunn} {et~al}\mbox{.}(2006){Gunn}, {Siegmund}, {Mannery}, {Owen},
  {Hull}, {Leger}, {Carey}, {Knapp}, {York}, {Boroski}, {Kent}, {Lupton},
  {Rockosi}, {Evans}, {Waddell}, {Anderson}, {Annis}, {Barentine}, {Bartoszek},
  {Bastian}, {Bracker}, {Brewington}, {Briegel}, {Brinkmann}, {Brown}, {Carr},
  {Czarapata}, {Drennan}, {Dombeck}, {Federwitz}, {Gillespie}, {Gonzales},
  {Hansen}, {Harvanek}, {Hayes}, {Jordan}, {Kinney}, {Klaene}, {Kleinman},
  {Kron}, {Kresinski}, {Lee}, {Limmongkol}, {Lindenmeyer}, {Long}, {Loomis},
  {McGehee}, {Mantsch}, {Neilsen}, {Neswold}, {Newman}, {Nitta}, {Peoples},
  {Pier}, {Prieto}, {Prosapio}, {Rivetta}, {Schneider}, {Snedden}, \&
  {Wang}}]{gun+06}
{Gunn} J.~E. {et~al.}, 2006, \aj, 131, 2332

\bibitem[{{Kalberla} {et~al}\mbox{.}(2005){Kalberla}, {Burton}, {Hartmann},
  {Arnal}, {Bajaja}, {Morras}, \& {P{\"o}ppel}}]{kal+05}
{Kalberla} P.~M.~W., {Burton} W.~B., {Hartmann} D., {Arnal} E.~M., {Bajaja} E.,
  {Morras} R., {P{\"o}ppel} W.~G.~L., 2005, \aap, 440, 775

\bibitem[{{Kirk} {et~al}\mbox{.}(1998){Kirk}, {Rieger}, \&
  {Mastichiadis}}]{kir+98}
{Kirk} J.~G., {Rieger} F.~M., {Mastichiadis} A., 1998, \aap, 333, 452

\bibitem[{{Komossa}(2002)}]{kom+02}
{Komossa} S., 2002, in Reviews in Modern Astronomy, Vol.~15, Reviews in Modern
  Astronomy, {Schielicke} R.~E., ed., p.~27

\bibitem[{{Komossa} \& {Bade}(1999)}]{kom+99}
{Komossa} S., {Bade} N., 1999, \aap, 343, 775

\bibitem[{{Komossa} \& {Dahlem}(2001)}]{kom+01}
{Komossa} S., {Dahlem} M., 2001, ArXiv Astrophysics e-prints

\bibitem[{{Maraschi} {et~al}\mbox{.}(1999){Maraschi}, {Fossati}, {Tavecchio},
  {Chiappetti}, {Celotti}, {Ghisellini}, {Grandi}, {Pian}, {Tagliaferri},
  {Treves}, {Breslin}, {Buckley}, {Carter-Lewis}, {Catanese}, {Cawley},
  {Fegan}, {Fegan}, {Finley}, {Gaidos}, {Hall}, {Hillas}, {Krennrich},
  {Lessard}, {Masterson}, {Moriarty}, {Quinn}, {Rose}, {Samuelson}, {Weekes},
  {Urry}, \& {Takahashi}}]{mar+99}
{Maraschi} L. {et~al.}, 1999, \apjl, 526, L81

\bibitem[{{Mason} {et~al}\mbox{.}(2001){Mason}, {Breeveld}, {Much}, {Carter},
  {Cordova}, {Cropper}, {Fordham}, {Huckle}, {Ho}, {Kawakami}, {Kennea},
  {Kennedy}, {Mittaz}, {Pandel}, {Priedhorsky}, {Sasseen}, {Shirey}, {Smith},
  \& {Vreux}}]{man+10}
{Mason} K.~O. {et~al.}, 2001, \aap, 365, L36

\bibitem[{{Oke} \& {Gunn}(1983)}]{oke+83}
{Oke} J.~B., {Gunn} J.~E., 1983, \apj, 266, 713

\bibitem[{{Pereira-Santaella} {et~al}\mbox{.}(2011){Pereira-Santaella},
  {Alonso-Herrero}, {Santos-Lleo}, {Colina}, {Jim{\'e}nez-Bail{\'o}n},
  {Longinotti}, {Rieke}, {Ward}, \& {Esquej}}]{san+11}
{Pereira-Santaella} M. {et~al.}, 2011, \aap, 535, A93

\bibitem[{{Persic} \& {Rephaeli}(2002)}]{per+02}
{Persic} M., {Rephaeli} Y., 2002, \aap, 382, 843

\bibitem[{{Rees}(1988)}]{rees88}
{Rees} M.~J., 1988, \nat, 333, 523

\bibitem[{{Saxton} {et~al}\mbox{.}(2012){Saxton}, {Read}, {Esquej}, {Komossa},
  {Dougherty}, {Rodriguez-Pascual}, \& {Barrado}}]{sax+12}
{Saxton} R.~D., {Read} A.~M., {Esquej} P., {Komossa} S., {Dougherty} S.,
  {Rodriguez-Pascual} P., {Barrado} D., 2012, \aap, 541, A106

\bibitem[{{Schlafly} \& {Finkbeiner}(2011)}]{sch+11}
{Schlafly} E.~F., {Finkbeiner} D.~P., 2011, \apj, 737, 103

\bibitem[{{Smith} {et~al}\mbox{.}(2001){Smith}, {Brickhouse}, {Liedahl}, \&
  {Raymond}}]{sim+01}
{Smith} R.~K., {Brickhouse} N.~S., {Liedahl} D.~A., {Raymond} J.~C., 2001,
  \apjl, 556, L91

\bibitem[{{Str{\"u}der} {et~al}\mbox{.}(2001){Str{\"u}der}, {Briel}, {Dennerl},
  {Hartmann}, {Kendziorra}, {Meidinger}, {Pfeffermann}, {Reppin}, {Aschenbach},
  {Bornemann}, {Br{\"a}uninger}, {Burkert}, {Elender}, {Freyberg}, {Haberl},
  {Hartner}, {Heuschmann}, {Hippmann}, {Kastelic}, {Kemmer}, {Kettenring},
  {Kink}, {Krause}, {M{\"u}ller}, {Oppitz}, {Pietsch}, {Popp}, {Predehl},
  {Read}, {Stephan}, {St{\"o}tter}, {Tr{\"u}mper}, {Holl}, {Kemmer}, {Soltau},
  {St{\"o}tter}, {Weber}, {Weichert}, {von Zanthier}, {Carathanassis}, {Lutz},
  {Richter}, {Solc}, {B{\"o}ttcher}, {Kuster}, {Staubert}, {Abbey}, {Holland},
  {Turner}, {Balasini}, {Bignami}, {La Palombara}, {Villa}, {Buttler},
  {Gianini}, {Lain{\'e}}, {Lumb}, \& {Dhez}}]{stu+01}
{Str{\"u}der} L. {et~al.}, 2001, \aap, 365, L18

\bibitem[{{Tu} \& {Wang}(2013)}]{tu+13}
{Tu} X., {Wang} Z.-X., 2013, Research in Astronomy and Astrophysics, 13, 323

\bibitem[{{Turner} {et~al}\mbox{.}(2001){Turner}, {Abbey}, {Arnaud},
  {Balasini}, {Barbera}, {Belsole}, {Bennie}, {Bernard}, {Bignami}, {Boer},
  {Briel}, {Butler}, {Cara}, {Chabaud}, {Cole}, {Collura}, {Conte}, {Cros},
  {Denby}, {Dhez}, {Di Coco}, {Dowson}, {Ferrando}, {Ghizzardi}, {Gianotti},
  {Goodall}, {Gretton}, {Griffiths}, {Hainaut}, {Hochedez}, {Holland},
  {Jourdain}, {Kendziorra}, {Lagostina}, {Laine}, {La Palombara}, {Lortholary},
  {Lumb}, {Marty}, {Molendi}, {Pigot}, {Poindron}, {Pounds}, {Reeves},
  {Reppin}, {Rothenflug}, {Salvetat}, {Sauvageot}, {Schmitt}, {Sembay},
  {Short}, {Spragg}, {Stephen}, {Str{\"u}der}, {Tiengo}, {Trifoglio},
  {Tr{\"u}mper}, {Vercellone}, {Vigroux}, {Villa}, {Ward}, {Whitehead}, \&
  {Zonca}}]{tur+01}
{Turner} M.~J.~L. {et~al.}, 2001, \aap, 365, L27

\bibitem[{{Wright} {et~al}\mbox{.}(2010){Wright}, {Eisenhardt}, {Mainzer},
  {Ressler}, {Cutri}, {Jarrett}, {Kirkpatrick}, {Padgett}, {McMillan},
  {Skrutskie}, {Stanford}, {Cohen}, {Walker}, {Mather}, {Leisawitz}, {Gautier},
  {McLean}, {Benford}, {Lonsdale}, {Blain}, {Mendez}, {Irace}, {Duval}, {Liu},
  {Royer}, {Heinrichsen}, {Howard}, {Shannon}, {Kendall}, {Walsh}, {Larsen},
  {Cardon}, {Schick}, {Schwalm}, {Abid}, {Fabinsky}, {Naes}, \&
  {Tsai}}]{wri+10}
{Wright} E.~L. {et~al.}, 2010, \aj, 140, 1868

\bibitem[{{Zhang} {et~al}\mbox{.}(2006){Zhang}, {Treves}, {Maraschi}, {Bai}, \&
  {Liu}}]{zhang+06}
{Zhang} Y.~H., {Treves} A., {Maraschi} L., {Bai} J.~M., {Liu} F.~K., 2006,
  \apj, 637, 699

\end{thebibliography}
